\documentclass{aastex63}

\newcommand{\Var}{\ensuremath{\mathbf{Var}\{}}
\newcommand{\beq}{\begin{equation}}
\newcommand{\eeq}{\end{equation}}

\definecolor{bgrn}{rgb}{0.0,0.5,0.5}
\definecolor{purp}{rgb}{0.75,0.5,0.5}

\usepackage{xcolor}
\usepackage{natbib}
\usepackage{amsmath, amsfonts, amssymb}
\usepackage{threeparttablex, longtable, booktabs}
\usepackage{appendix}

\newcounter{alignfirst}

\submitjournal{AJ}

\begin{document}

\title{Magnitude-squared coherence: A powerful tool for disentangling Doppler planet discoveries from stellar activity}

\correspondingauthor{Sarah Dodson-Robinson}
\email{sdr@udel.edu}

\author[0000-0002-8796-4974]{Sarah E. Dodson-Robinson}
\affiliation{University of Delaware \\
Bartol Research Institute \\
Department of Physics and Astronomy \\
Newark, DE 19716, USA}

\author[0000-0001-8183-459X]{Victor Ramirez Delgado}
\affiliation{University of Delaware \\
Department of Physics and Astronomy \\
Newark, DE 19716, USA}

\author[0000-0001-6771-4583]{Justin Harrell}
\affiliation{University of Delaware \\
Department of Physics and Astronomy \\
Newark, DE 19716, USA}

\author[0000-0003-3996-773X]{Charlotte L. Haley}
\affiliation{Argonnne National Laboratory \\
Mathematics and Computer Science Division \\
9700 S Cass Ave \\
Lemont, IL 60439, USA}

\begin{abstract}

If Doppler searches for earth-mass, habitable planets are to succeed, observers must be able to identify and model out stellar activity signals. Here we demonstrate how to diagnose activity signals by calculating the magnitude-squared coherence $\hat{C}^2_{xy}(f)$ between an activity indicator time series $x_t$ and the radial velocity (RV) time series $y_t$. Since planets only cause modulation in RV, not in activity indicators, a high value of $\hat{C}^2_{xy}(f)$ indicates that the signal at frequency $f$ has a stellar origin. We use Welch's method to measure coherence between activity indicators and RVs in archival observations of GJ~581, $\alpha$~Cen~B, and GJ~3998. High RV-H$\alpha$ coherence at the frequency of GJ~3998 b, and high RV-S index coherence at the frequency of GJ~3998 c, indicate that the planets may actually be stellar signals. We also replicate previous results showing that GJ~581 d and g are rotation harmonics and demonstrate that $\alpha$~Cen~B has activity signals that are not associated with rotation. Welch's power spectrum estimates have cleaner spectral windows than Lomb-Scargle periodograms, improving our ability to estimate rotation periods. We find that the rotation period of GJ~581 is 132 days, with no evidence of differential rotation. Welch's method may yield unacceptably large bias for datasets with $N < 75$ observations and works best on datasets with $N > 100$. Tapering the time-domain data can reduce the bias of the Welch's power spectrum estimator, but observers should not apply tapers to datasets with extremely uneven observing cadence. A software package for calculating magnitude-squared coherence and Welch's power spectrum estimates is available on github. 






\end{abstract}

\keywords{Time series analysis (1916), Period search (1955), Astrostatistics techniques (1886), Radial velocity (1332), Stellar activity (1580), Stellar rotation (1629)}


\section{Introduction}
\label{sec:intro}

Planets are often diagnosed in radial velocity (RV) periodograms as large peaks which register above some high significance level. While some periodogram peaks genuinely describe planets, others are spurious detections resulting from
stellar rotation or activity \citep[e.g.][]{saar97, hatzes02, desort07, boisse11, robertson14a, robertson14b, newton16, suarezmascareno17b, rajpaul21}. Now that extreme precision spectrographs are on the hunt for terrestrial planets \citep{jurgenson16, gonzalezhernandez18, gupta21}, the need for high-performance activity diagnostics is urgent: while an earthlike planet orbiting a sunlike star yields an RV oscillation with semiamplitude $K < 10$~cm~s$^{-1}$, that same star's expected rotational RV modulation has amplitude $> 1$~m~s$^{-1}$ \citep{vanderburg16}. Doppler surveys won't yield any Earth analogs unless stellar signals can be accurately identified and modeled.

A sure sign that a peak in the RV periodogram is caused by stellar activity is when the periodogram of a simultaneously measured activity indicator such as H$\alpha$, Mt.\ Wilson S-index, or bisector span shows the same peak (or its harmonic) \citep[e.g.][]{queloz01, bonfils07, kane16, sarkis18, toledopadron19}. However, for quiet target stars with low-level variability, the signal in the activity-indicator power spectrum might not hit a statistically significant false alarm probability. In such cases, the connection between stellar activity and RV might be overlooked \citep[e.g.][]{robertson15, bortle21, lubin21}. Furthermore, aliasing, small sample size, and red noise can cause bootstrap false alarm level calculations to fail \citep{baluev08, chernick08, littlefair16}, making it difficult to correctly assess the significance of activity signals. Another common diagnostic of stellar activity---a linear regression of RV onto an activity indicator \citep[e.g.][]{queloz01, huelamo08, queloz09, talor18}---fails when the RV and activity signals are not in phase, as seen in the RV, S-index, and H$\alpha$-index measurements of 55~Cnc \citep{butler17, bourrier18}. To validate planet discoveries, we require analysis techniques that reveal all oscillatory components common to simultaneous time series, regardless of relative phase.

In this paper, we demonstrate the use of the magnitude-squared coherence in RV planet searches. This bivariate statistic diagnoses oscillations that manifest in more than one observable. Magnitude-squared coherence can be interpreted as a frequency-dependent correlation coefficient that describes the proportion of variance in one time series that can be explained by means of a lagged linear regression onto another time series. For example, if RV and and $\log R^{\prime}_{HK}$ \citep{noyes84} both oscillate at the star rotation frequency---as happened when $\alpha$~Cen~B had a large spot group in 2010 \citep{dumusque12, thompson17}---the coherence would be high at that frequency, and it would be possible to predict RV using a lagged regression onto $\log R^{\prime}_{HK}$ {\it if the observations were regularly spaced in time} \citep{shumwaystoffer}. Bivariate statistics are used in a myriad of physical science applications, such as solar physics \citep{walker14}, climatology \citep{thomson95}, oceanography \citep{chave92, miller21}, atmospheric science \citep{krug19}, seismology \citep{scafetta15}, and others \citep{carter87}.

This paper is organized as follows. In \S \ref{sec:math}, we present the mathematical fundamentals of the magnitude-squared coherence $C^2_{xy}(f)$ and its two companion bivariate statistics, the cross-spectrum and phase spectrum. In \S \ref{sec:computation}, we describe computational methods used to estimate $C^2_{xy}(f)$. In \S \ref{sec:application}, we demonstrate our methods' stellar activity diagnostic power using published RV observations of $\alpha$~Cen~B, GJ~581, and GJ~3998. We present our conclusions and plans for future work in \S \ref{sec:conclusions}. Appendix \ref{sec:bivariate} introduces our new publicly available software package, \texttt{NWelch} \citep{dodsonrobinson22}, which was used for all calculations involving real RV data that are presented here. 




\begin{deluxetable}{ll}
\tablecaption{Definitions of symbols}
\tablehead{
\colhead{Symbol} & \colhead{Definition}
}
\startdata
$t_i$                         & time stamp \\
$\Delta t_i$                  & time between observations \\
$x_t$, $y_t$                  & time series \\
$N$                           & number of data points in time series \\
$\mu_{x}$                     & expected value of series $x_t$ \\
$\bar{x}$                     & sample mean of series $x_t$ \\
$E\{ \cdot \}$                & expected value \\
$\gamma_{xy}(\tau)$           & cross-correlation between series $x_t$ and $y_t$ \\
$\hat{\gamma}_{xy}(\tau)$     & sample cross-correlation between series $x_t$ and $y_t$ \\
$\tau$                        & time lag (independent variable in cross-correlation) \\
$\mathcal{F} \{ \}$           & (Nonuniform) Fourier transform \\
$S_{xy}(f)$                   & cross-spectrum of time series $x_t$ and $y_t$ \\ 
$\hat{S}_{xy}(f)$             & estimated cross-spectrum of time series $x_t$ and $y_t$ \\ 
$\hat{c}_{xy}(f)$             & estimated cospectrum between time series $x_t$ and $y_t$ \\
$\hat{q}_{xy}(f)$             & estimated quadrature spectrum between time series $x_t$ and $y_t$ \\
$\hat{C}^2_{xy}(f)$           & estimated magnitude-squared coherence between $x_t$ and $y_t$ \\
$\hat{S}_{xx}(f)$, $\hat{S}_{yy}(f)$ & estimated power spectra of time series $x_t$ and $y_t$ \\
$\hat{\phi}_{xy}$             & estimated relative phase spectrum of $x_t$ and $y_t$ \\
$*$                           & convolution \\
$\hat{a}$                     & estimated value of $a$ \\
$\Var \cdot \}$               & variance \\
$z(f)$                        & Fisher's variance-stabilizing transformation of $\hat{C}^2_{xy}(f)$ \\
$w_t$ & taper coefficients \\
$W(f)$                          & spectral window \\
$K$                           & number of segments used in Welch's algorithm \\
$\widetilde{K}$               & effective number of independent estimates $\hat{S}_{xx}(f)$, $\hat{S}_{yy}(f)$, $\hat{S}_{xy}(f)$ \\
$x^{(k)}_j$, $y^{(k)}_j$, $w_j^{(k)}$ & observations and taper coefficients belonging to segment $k$ \\
$\hat{S}_{xx}^{(k)}(f)$, $\hat{S}_{yy}^{(k)}(f)$, $\hat{C}_{xy}^{(k)}(f)$ & $S_{xx}(f)$, $S_{yy}(f)$, $S_{xy}(f)$ estimated using $x_t^{(k)}$, $y_t^{(k)}$ taken from segment $k$ \\
$\hat{S}^w_{xx}(f)$, $\hat{S}^w_{yy}(f)$, $\hat{S}^w_{xy}(f)$
                                & tapered estimates of $S_{xx}(f)$, $S_{yy}(f)$ $S_{xy}(f)$ \\
$N^{(k)}$                       & number of data points in segment $k$ \\
$\mathcal{R}^{(k)}$             & Rayleigh resolution of $\hat{S}_{xy}^{(k)}(f)$\\
${\rm bias} \left[ \hat{C}^2_{xy}(f) \right]$ & approximate bias of magnitude-squared coherence estimator \\
$\hat{C}^{\prime \; 2}_{xy}(f)$  & debiased magnitude-squared coherence estimate \\
FAL                             & false-alarm level \\
$\alpha$                        & false-alarm probability \\
$\mathcal{B}$                   & half-width of the main lobe of the spectral window \\
$\mathbf{max}[\cdot]$           & maximum value \\
$g_r$                           & threshold periodogram power for Fisher's test \\
$g$                             & threshold periodogram power for Siegel's test \\
$N_f$                           & number of entries in the frequency grid \\
$\lambda$                       & proportionality constant relating $g$ and $g_r$ \\
$T_{\lambda}$                   & Siegel's test statistic \\
\enddata
\end{deluxetable}

\section{Mathematical fundamentals of magnitude-squared coherence}
\label{sec:math}

Suppose a planet-search team is lucky enough to have an extreme-precision spectrograph in space. There are no equipment problems that require telescope downtime, and all data downlinks can be completed in under 24 hours. The team has the luxury of observing a sunlike target at the same time every Earth day, with no interruptions, for many years. The resulting observations are evenly spaced, with constant $\Delta t = t_{i+1} - t_i = 1$~day (where $t$ is a time stamp and $i$ is an integer index; symbol definitions are collected in Table 1). From each astronomical spectrum, the data pipeline returns RV plus dozens of activity indicators \citep[e.g.][]{wise18}. The team has assembled the perfect multivariate planet-search time series, which should allow members to characterize all stellar signals.

\setcounter{alignfirst}{3}

\begin{figure}
    \centering
    \includegraphics[width=0.8\textwidth]{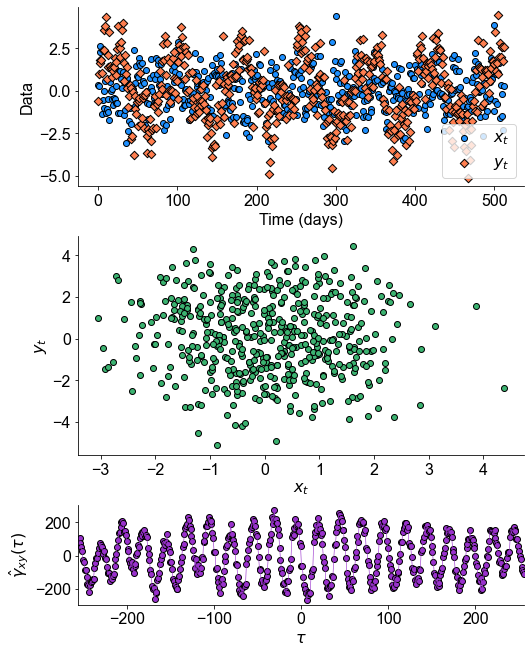}
    \caption{{\bf Top:} Synthetic time series $x_t$ (Equation \thealignfirst), which represents an activity indicator, and $y_t$ (Equation \ref{eq:examplexy}), which represents radial velocity. Star rotation manifests in both $x_t$ and $y_t$, and $y_t$ also records planet-induced oscillations. {\bf Middle:} A plot of $y_t$ as a function of $x_t$ fails to reveal a straight-line relationship between the two observables. This is because (1) the rotation signal in $y_t$ lags the rotation signal in $x_t$ by $90^{\circ}$, and (2) both time series are plagued with Gaussian white noise. {\bf Bottom:} The cross-correlation 
    $\hat{\gamma}_{xy}(\tau)$ shows an oscillation at the rotation period.}
    \label{fig:example1}
\end{figure}

Knowing that any signal manifesting in both an activity indicator and RV time series cannot be driven by an earthlike planet \citep[after all, exo-Earths are too miniscule and too far from their host stars for tides to trigger stellar activity as in HD~179949;][]{shkolnik03}, the team decides to use the cross-correlation as an activity diagnostic. 
The cross-correlation between two jointly stationary time series $x_t$ and $y_t$ is defined as 
\begin{equation}
\gamma_{xy}(\tau) = E\{(x_t - \mu_x)(y_{t+\tau} - \mu_y)\}
\label{eq:xcor}
\end{equation}
where $E\{ \cdot \}$ denotes expected value, $\mu_{x} = E\{x_t\}$, $\mu_y = E\{y_t\}$, and $\tau$ is an integer time lag. When observed at $N$ contiguous time points $t = 0, \ldots, N-1$, $\gamma_{xy}(\tau)$ can be estimated using the convolution formula \cite[Equations (1.29), (1.40)]{shumwaystoffer}
\begin{equation}
    \hat{\gamma}_{xy}(\tau) = \frac{1}{N} \sum_{t = 0}^{N - \tau - 1}  (x_t - \bar{x})(y_{t+\tau} - \bar{y}), \label{eq:crosscorrelation}
\end{equation}
where $\bar{x} = N^{-1} \sum_{t = 0}^{N-1} x_t$ is the sample mean of the series $x_t$, $\bar{y}$ is the sample mean of the series $y_t$, and $\hat{a}$ denotes an estimate of $a$.
Oscillations with period $P$ that manifest in both $x_t$ and $y_t$ will yield local maxima in 
$\hat{\gamma}_{xy}(\tau)$ at integer multiples of $P$ and local minima where the lag is such that the oscillation is perfectly out of phase, generating a wave pattern that is easy to see in a plot. The top panel of Figure \ref{fig:example1} shows two synthetic regularly spaced time series with $N = 512$ representing possible spacecraft measurements. Let $x_t$ denote a noisy activity indicator time series that records the star rotation with period $P_{\rm rot} = 25$~days, while $y_t$ is the noisy RV time series that shows both the rotation and a planet with period $P_{\rm pl} = 80$~days:
\begin{align}
x_t &= \cos(2 \pi t f_{\rm rot}) + \zeta_t \\ 
y_t &= \sin(2 \pi t f_{\rm rot}) + 2 \sin (2 \pi t f_pl) + \eta_t, 
\label{eq:examplexy}
\end{align}
where $f_{\rm rot} = 1/P_{\rm rot}$ is the rotation frequency, $f_{\rm pl} = 1/P_{\rm pl}$ is the planet frequency, and 
$\zeta_t$ and $\eta_t$ denote uncorrelated Gaussian white noise processes with zero mean and unit variance. The middle panel of Figure \ref{fig:example1}, which shows $y_t$ as a function of $x_t$, demonstrates that there is no zero-lag, straight-line relationship between the two observables because of both the phase shift in the shared rotation signal and the noise in each time series. But the shared oscillation is obvious in the bottom panel of Figure \ref{fig:example1}, which reveals a wave pattern with period $P_{\rm rot}$ in 
$\hat{\gamma}_{xy}(\tau)$. 

\begin{figure}
    \centering
    \includegraphics[width=0.8\textwidth]{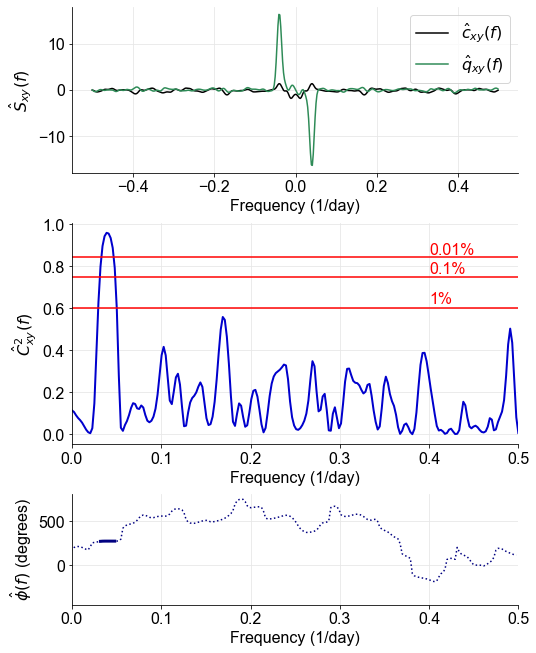}
    \caption{{\bf Top:} The cross-spectrum estimate $\hat{S}_{xy}(f)$
    of synthetic signals $x_t$ and $y_t$ (Equation \ref{eq:crossspectrum}). Spikes at $\pm f_{\rm rot}$ appear in $\hat{q}_{xy}(f)$ because the rotation signal in $y_t$ lags the signal in $x_t$. {\bf Middle:} Estimated magnitude-squared coherence between $x_t$ and $y_t$ (Equation \ref{eq:coherence}). From bottom to top, the red horizontal lines show the 1\%, 0.1\%, and 0.01\% false alarm levels. The shared rotation signal generates a strong peak in $\hat{C}^2_{xy}(f)$, but the planet signal present only in $y_t$ does not show up in $\hat{C}^2_{xy}(f)$. {\bf Bottom}: Estimated phase spectrum $\hat{\phi}_{xy}(f)$. The dotted line shows the estimated phase at all frequencies. The solid line shows the  part of the phase estimate at frequencies near $f_{\rm rot}$, where the coherence is statistically significant. As expected from Equations \thealignfirst\ and \ref{eq:examplexy}, $\hat{\phi}_{xy}(f_{\rm rot}) = 270^{\circ}$. 
    }
    \label{fig:example2}
\end{figure}

If the real data are noisier than our synthetic $x_t$ and $y_t$ or the two time series share more than one oscillation---such as rotation and a long-term activity cycle---it is more difficult to identify the periods of shared signals in a plot of 
$\hat{\gamma}_{xy}(\tau)$. To pinpoint the frequencies of oscillations in common to both $x_t$ and $y_t$, the planet-search team can 
estimate the (complex valued) frequency-dependent cross-correlation,
or \emph{cross-spectrum}, using
\begin{align}
    S_{xy}(f) &= \int_{-\infty}^{\infty} \gamma_{xy}(\tau) e^{-2\pi i f \tau} d\tau \\
    \hat{S}_{xy}(f) &= \frac{1}{N} \sum_{t = 0}^{N-1} (x_t - \bar{x}) e^{-2\pi i f t}  \sum_{s = 0}^{N-1} (y_s - \bar{y}) e^{2\pi i f s},
\label{eq:crossspectrum}
\end{align}
where $f$ is the frequency.
In-phase and $180^{\circ}$ out-of-phase 
sinusoids that occur in both $x_t$ and $y_t$ yield delta functions in  
$c_{xy}(f) = \operatorname{Re} \{ S_{xy}(f) \}$ (the cospectrum)\footnote{Actually the signals that appear in the cross-spectrum will be shaped by the spectral window (\S \ref{subsec:tapering}), but for this ``perfect dataset'' example they will approximate delta functions in the limit as $N\rightarrow \infty$.}, while $\pm 90^{\circ}$ phase-shifted oscillations in common to $x_t$ and $y_t$ show up as delta functions in $q_{xy}(f) = \operatorname{Im} \{ S_{xy}(f) \}$ (the quadrature spectrum). In general, the cospectrum---an even function---is large when there is a small phase angle near zero, and the quadrature spectrum---an odd function---is large and positive when there is a phase angle near $90^{\circ}$, and will be negative 
when the phase separation approaches 270$^{\circ}$. Coherent oscillations with relative phases that are not integer multiples of $90^{\circ}$ will yield delta functions in both the cospectrum and the quadrature spectrum. Where $x_t$ and $y_t$ share multiple oscillations, the cross-spectrum will have delta functions at all oscillation frequencies. The top panel of Figure \ref{fig:example2} shows $\hat{S}_{xy}(f)$ for our synthetic spacecraft dataset. The shared but phase-shifted rotation signal creates spikes in the quadrature spectrum at $f = \pm 1/25$~days$^{-1}$.

One can normalize $S_{xy}(f)$ to produce the frequency-dependent cross-correlation coefficient, or magnitude squared coherence
\begin{align}
    C^2_{xy}(f) &= \frac{|S_{xy}(f)|^2}{S_{xx}(f) S_{yy}(f)} \\
    \hat{C}^2_{xy}(f) &=  \frac{|\hat{S}_{xy}(f)|^2}{\hat{S}_{xx}(f) \hat{S}_{yy}(f)}.
\label{eq:coherence}
\end{align}
To compute power spectrum estimates $\hat{S}_{xx}(f)$ and $\hat{S}_{yy}(t)$, replace the $(x_t - \bar{x})(y_t - \bar{y})$ in Equation \ref{eq:crossspectrum} with $(x_t - \bar{x})(x_s - \bar{x})$ or $(y_t - \bar{y})(y_s - \bar{y})$. 
In Equation \ref{eq:coherence}, $\hat{S}_{xx}(f)$ and $\hat{S}_{yy}(t)$ are normalization factors that control for the fact that different observables might have vastly different amplitude variations---for example, RV semiamplitudes are of order unity or larger even for quiet sunlike stars \citep{vanderburg16}, while the H$\alpha$ index defined by \citet{gomesdasilva11} records oscillations at the 1\% level \citep[e.g.][]{robertson15}. If $x_t$ and $y_t$ are activity indicator and RV, respectively, $\hat{C}^2_{xy}(f)$ should be near zero at the orbital frequency of any planet candidate. 
As $\hat{C}^2_{xy}(f)$ approaches unity, it lends more support to the hypothesis that the velocity signal at $f$ is caused by stellar activity. 
The estimated phase spectrum, or frequency-dependent phase lag between the two time series, is
\begin{equation}
    \hat{\phi}_{xy}(f) = \arctan \left( \frac{\hat{c}_{xy}(f)}{\hat{q}_{xy}(f)} \right).
\label{eq:phase}
\end{equation}
$\hat{\phi}_{xy}(f)$ is especially useful at frequencies where $\hat{C}^2_{xy}(f)$ exceeds some threshold of statistical significance.
The middle panel of Figure \ref{fig:example2} shows the estimated magnitude-squared coherence $\hat{C}^2_{xy}(f)$
between our example $x_t$ and $y_t$, while the bottom panel shows the estimated phase spectrum $\hat{\phi}_{xy}(f)$.
The shared rotation signal shows up in $\hat{C}^2_{xy}(f)$ as a strong peak at $f_{\rm rot}$, while the planet signal at $f_{\rm pl}$ that appears only in RV does not show up in $\hat{C}^2_{xy}(f)$ at all. This example shows the power of magnitude-squared coherence in separating activity signals from planets.\footnote{For our synthetic dataset, $\hat{S}_{xy}(f)$ and $\hat{C}^2_{xy}(f)$ were computed in \texttt{python 3} with \texttt{scipy.signal.csd} and \texttt{scipy.signal.coherence}, respectively. To estimate $\hat{C}^2_{xy}(f)$, each time series was divided into six overlapping segments and a Blackman-Harris taper was applied to each segment. See \S \S \ref{subsec:segmenting} and \ref{subsec:tapering} for more on segmenting and tapering the time series.}

Of course, we do not have an extreme-precision spectrograph on a spacecraft with no-fail instrumentation and rapid downlink. Like all ground-based astronomical time series, RV and activity-indicator datasets are sampled at uneven intervals, with daytime and seasonal gaps. When the time between spectroscopic observations $\Delta t$ is not constant, no direct cross-correlation estimate 
$\hat{\gamma}_{xy}(\tau)$ can be calculated. But with a non-uniform fast Fourier transform algorithm, it is still possible to estimate the cross-spectrum $\hat{S}_{xy}(f)$ between RV and an activity indicator \citep{scargle89}. 
Define the nonuniform fast Fourier transform (NFFT) of the sequence $x_{j}$ observed on the sequence of times $t_j$ where $j = 0, \ldots, N-1$ implicitly as
\begin{equation} \label{eq:NFFT}
x_j = \sum_{k = 0}^{N-1} \tilde{x}(f_k) e^{i 2\pi t_j f_k},
\end{equation}
where $k$ is the index of the frequency grid and 
$t_j$ has been standardized to the scaled time interval $[-1/2, 1/2)$. That is, to solve for the coefficients $\tilde{x}(k)$ one computes $(\boldmath{A}^*)^T$, the \emph{adjoint} of the matrix $\boldmath{A}$ with entries $\boldmath{A}_{jk} = e^{2\pi i f_k t_j}$ \citep{keiner2009using}, and multiplies $(\boldmath{A}^*)^T$ by the column vector of observations $x_{j}$ \citep{springford2020improving}. (See \S \ref{app:NUFFT} for more on the NFFT algorithm.) In what follows, we use the notation $\tilde{x}(f)$ to denote the NFFT on a grid of equally spaced Fourier frequencies $f_k$.
When the data are equally spaced in the time domain, the inverse transform can be easily written in closed form. Replacing the Fourier transform with NFFTs, one can obtain the following (normalized) power spectral density estimator
\begin{equation}
    \hat{S}_{xx}(f) = |\tilde{x}(f)|^2
\end{equation}
and similarly, an estimator for the cross spectrum 
\begin{equation}
    \hat{S}_{xy}(f) = \tilde{x}(f) \tilde{y}^*(f)
    \label{eq:naiveSxy}
\end{equation}
(where $*$ deontes complex conjugate), which has the desired property that convolution of the sequences $x_t$ and $y_t$ results in the multiplication of their Fourier transforms, i.e. $\tilde{x}(f) \tilde{y}(f)$. For a description of the NFFT algorithm, see \S \ref{app:NUFFT}.


Equation \ref{eq:naiveSxy} reveals a trap for the unwary: if the magnitude-squared coherence is calculated based only on a {\it single} estimate of each of $\hat{S}_{xx}(f)$, $\hat{S}_{yy}(f)$, and $\hat{S}_{xy}(f)$, the disastrous result will be $\hat{C}^2_{xy}(f) = 1$. To see why, we substitute Equation \ref{eq:naiveSxy} into Equation \ref{eq:coherence}: the result is $S_{xx}(f) S_{yy}(f) / [S_{xx}(f) S_{yy}(f)]$. But if we have $K >1$ estimates of $\hat{S}_{xy}(f)$, say $\hat{S}^{(k)}_{xy}(f)$ for $k = 0, \ldots, K-1$, we can take advantage of the fact that 
\begin{equation}
   \left | \sum_{k = 0}^{K-1} \hat{S}^{(k)}_{xy}(f) \right|^2 \neq \sum_{k = 0}^{K-1} | \hat{S}^{(k)}_{xy}(f)|^2.
\label{eq:averagingestimates}
\end{equation}
Thus a meaningful coherence estimate comes from averaging together multiple estimates of the numerator and denominator of Equation \ref{eq:coherence} before computing their ratio: 
\begin{equation}
    \hat{C}^2_{xy}(f) = \frac{ | \sum_{k = 0}^{K-1} \hat{S}^{(k)}_{xy}(f) |^2}{ \sum_{k = 0}^{K-1} \hat{S}^{(k)}_{xx}(f) \; \sum_{k = 0}^{K-1} \hat{S}^{(k)}_{yy}(f)}.
\label{eq:estcoh}
\end{equation}
The question becomes, how do we obtain the $\hat{S}^{(k)}$? For RV datasets, we will divide each time series into shorter segments and compute one estimate of each of $\hat{S}_{xx}(f)$, $\hat{S}_{yy}(f)$, and $\hat{S}_{xy}(f)$ from each segment (\S \ref{subsec:segmenting}). The signal processing literature describes this procedure as Welch's method.

\begin{figure}
    \centering
    \includegraphics[width=0.6\textwidth]{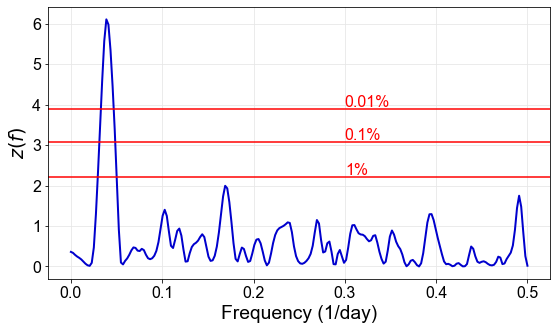}
    \caption{$z(f)$, the Fisher variance-stabilizing $\operatorname{atanh}$ transformation of $\hat{C}^2_{xy}(f)$. 1\%, 0.1\%, and 0.01\% false alarm levels are shown in red.}
    \label{fig:transformed}
\end{figure}

For bivariate frequency-domain analysis, there is one more useful mathematical operation. As mentioned above, 
$C^2_{xy}(f)$ is bounded between 0 and 1, which means $\Var \hat{C}^2_{xy}(f) \}$ is a function of $\hat{C}^2_{xy}(f)$ (where $\Var \cdot \}$ denotes variance). In practical terms, this means the difference between (for example) $\hat{C}^2_{xy}(f) = 0.91$ and $\hat{C}^2_{xy}(f) = 0.92$ might be more statistically significant than the difference between $\hat{C}^2_{xy}(f) = 0.5$ and $\hat{C}^2_{xy}(f) = 0.6$ (depending on $N$ and $K$). To stabilize $\Var \hat{C}^2_{xy}(f) \}$ and remove some of its dependence on $\hat{C}^2_{xy}(f)$, we can use Fisher's $z$ transformation \citep{fisher29, jenkins68}:
\begin{equation}
    z(f) = \sqrt{2 K - 2} \operatorname{atanh} \left[ \hat{C}^2_{xy}(f) \right],
\end{equation}
where $\operatorname{atanh}$ is the inverse hyperbolic arctangent. 
The transformed coherence $z(f)$ is approximately Student$-t$ distributed \citep{tc91}. 
Figure \ref{fig:transformed} shows $z(f)$ from the $\hat{C}^2_{xy}(f)$ estimate plotted in Figure \ref{fig:example2}. 

In the next section, we describe the computational methods used to estimate $\hat{S}_{xy}(f)$, $\hat{C}^2_{xy}(f)$, and $\hat{\phi}_{xy}(f)$ for RV data.

\section{Computational methods}
\label{sec:computation}

The biggest challenge in applying Equation \ref{eq:estcoh} is obtaining multiple estimates of $\hat{S}_{xy}(f)$. For evenly spaced data (constant $\Delta t$), there are three possibilities: the multitaper method of \citet{t82}, where the time-domain data are multiplied by a set of orthogonal sequences \citep{s78} and the results from subsequent Fourier analysis are averaged together via jackknife mean \citep[see a bivariate application of this technique in][]{thomson95}, smoothing the cross spectrum across frequency \citep{shumwaystoffer}, or Welch's method \citep{welch67}.\footnote{A forthcoming paper will extend multitaper analysis to RV data (Dodson-Robinson et al.\ in preparation). That paper will build on the work of \citet{springford2020improving}, who applied the multitaper technique to {\it Kepler} data with near-constant $\Delta t$.} Here we demonstrate how to apply the computationally lightweight Welch's method to RV data. Our methods borrow heavily from the software package \texttt{redfit-x} \citep{olafsdottir16} and its predecessor \texttt{SPECTRUM} \citep{schulz97}, which implement Welch's method for analysis of unevenly spaced, bivariate paleoclimate data.

\subsection{Segmenting the data}
\label{subsec:segmenting}

\begin{figure}
    \centering
    \includegraphics[width=0.8\textwidth]{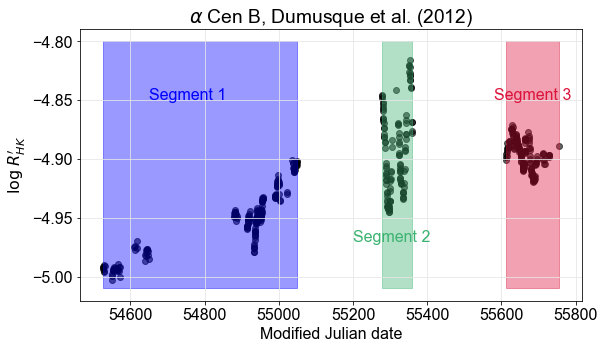}
    \caption{Non-overlapping segmenting scheme for the $\alpha$~Cen~B spectroscopic dataset from \citet{dumusque12}.}
    \label{fig:aCenB_segments}
\end{figure}

\citet{welch67} built on the work of \citet{bartlett48}, who proposed estimating the power spectrum of a stationary, regularly spaced time series $x_t$ by dividing the series into segments, computing a periodogram of each segment, and averaging the periodograms:
\begin{equation}
   \overline{S}_{xx}(f) = \frac{1}{K} \sum_{k = 0}^{K-1} \hat{S}_{xx}^{(k)}(f) = \frac{1}{K} \sum_{k = 0}^{K-1} \frac{|\tilde{x}^{(k)}(f) |^2}{N^{(k)}},
   \label{eq:segments}
\end{equation}
where $K$ is the number of segments
and $\tilde{x}^{(k)}(f)$ denotes the NFFT of the data subsequence $x^{(k)}_j$ of length $N^{(k)}$.\footnote{In Welch's original work on regularly spaced time series, $\tilde{x}^{(k)}(f)$ is the fast Fourier transform and $N^{(k)} = N^{(j)}$ for all $j, k = 0, \ldots, K-1$.} For bivariate time series observed at the same time stamps $t_j$, one uses the same segmentation scheme for both $x_t$ and $y_t$ (i.e.\ two observables measured from the same astronomical spectrum taken at time $t_j$ are assigned to the same segment $k$) and computes 
$\hat{S}_{xy}^{(k)}(f)$, $\hat{S}_{xx}^{(k)}(f)$, and $\hat{S}_{yy}^{(k)}(f)$ for each segment. Averaging together the $K$ different estimates of the cross-spectrum and power spectra yields  $\overline{S}_{xy}(f)$, $\overline{S}_{xx}(f)$, and $\overline{S}_{yy}(f)$---all the ingredients needed to compute $\hat{C}^2_{xy}$ using Equation \ref{eq:estcoh}. Figure \ref{fig:aCenB_segments} shows the \citet{dumusque12} $\alpha$~Cen~B $\log  R^{\prime}_{HK}$ time series divided into non-overlapping segments in preparation for using Equation \ref{eq:segments}.

\citet{bartlett48} and \citet{welch67} explored time-series segmenting not because they wanted to compute magnitude-squared coherences, but because they were looking for a power spectrum estimator whose variance decreases as the number of observations increases. This is not true of either the standard periodogram \citep{schuster1898} or the Lomb-Scargle periodogram \citep{lomb76, scargle82}: for both estimators, the variance is independent of the number of observations. \citet{welch67} pointed out that non-overlapping segments (aka Bartlett's method) are ideal for reducing the variance of any $\hat{S}(f)$, as the various $\hat{S}^{(k)}(f)$ are fully independent of one another.
But there is a catch. Although both the standard periodogram and the Lomb-Scargle periodogram are asymptotically unbiased---i.e.\ $\hat{S}(f) \rightarrow S(f)$ as $N \rightarrow \infty$---the bias can be severe for small or even not-so-small $N$ \citep[``tragedy of the periodogram,''][]{percival94}. To understand bias, suppose an RV time series $y_t$ traces a planet in a circular orbit with period $P_{pl}$. Since the planet's time domain signature is a perfect sinusoid, $\hat{S}_{yy}(f)$ should have an infinitely thin delta function at $f_{pl} = 1 / P_{pl}$. But the finite duration of $y_t$ creates spectral leakage:
in $\hat{S}_{yy}(f)$, the planet's signal will land mostly on the frequency $f_j$ nearest to $f_{pl}$, but there will be non-zero contributions to {\it every} frequency in the grid \citep[e.g.][]{harris78}. This leakage is responsible for periodogram bias. 
Thus there is a tradeoff between bias and variance: we want a large number of segments $K$ in order to improve the consistency of $\hat{S}_{xx}(f)$, $\hat{S}_{yy}(f)$, and $\hat{C}^2_{xy}(f)$, but if $N^{(k)}$ is too low our power spectrum and coherence estimates will be consistently biased---that is, $\hat{C}_{xy}(f)$ will be far away from $C_{xy}(f)$---especially at high frequencies \citep[e.g.][]{podesta06, bronez92, percival93}. 

\subsection{Overlapping segments}
\label{subsec:overlap}

\begin{figure}
    \centering
    \begin{tabular}{cc}
    \includegraphics[width=0.49\textwidth]{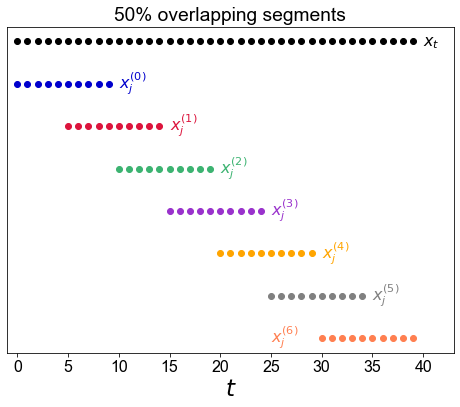} &
    \includegraphics[width=0.49\textwidth]{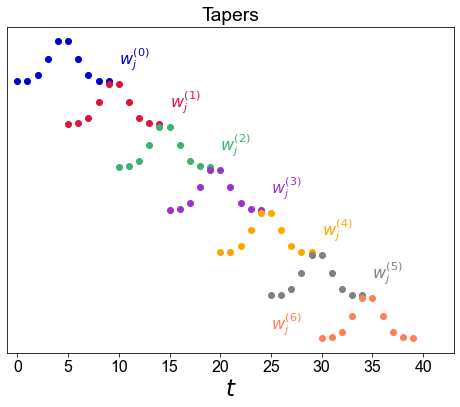} \\
    \end{tabular}
    \caption{{\bf Left:} Cartoon showing the allocation of data points in $x_t$ among segments $0 \ldots K-1$ in Welch's method. Series $x_t$ has 40 data points, which are broken into seven overlapping segments $x_j^{(0)}, \ldots, x_j^{(6)}$ that each have 10 data points. Note that 10 data points per segment are far too few to compute any kind of periodogram, let alone a minimally biased one; this figure uses small $N^{(k)}$ simply for the sake of readability. {\bf Right:} Tapers $w_j^{(k)}$ applied to each segment shown in the left panel. }
    \label{fig:overlap}
\end{figure}

To reduce the bias of each $\hat{S}^{(k)}(f)$ while retaining most of the variance suppression associated with large $K$, \citet{welch67} proposed using tapered, overlapping segments (tapering is discussed in \S \ref{subsec:tapering}). 
In the 50\% overlap scheme, we segment $x_t$ as follows:
\begin{align*}
    x^{(0)}_j &= x_0 \ldots x_{N^{(k)}} \\
    x^{(1)}_j &= x_{N^{(k)}/2} \ldots x_{3N^{(k)}/2 - 1} \\
    x^{(2)}_j &= x_{N^{(k)}} \ldots x_{2N^{(k)} - 1} \\
    x^{(3)}_j &= x_{3N^{(k)}/2} \ldots x_{5N^{(k)}/2 - 1} \\
    & \ldots \\
    x^{(K-1)}_j &= x_{N-N^{(k)}} \ldots x_{N-1}. \\
    \label{eq:welchoverlap}
\end{align*}
Each 50\% overlapping segment has $2 N / (K+1)$ data points, as opposed to $N/K$ for non-overlapping segments. But because of the overlap, the resulting spectral estimates $\hat{S}^{(k)}(f)$ are not independent. The variance reduction associated with overlapping segments is 
\begin{equation}
\Var \bar{S}(f) \} = \frac{1}{\widetilde{K}} \Var \hat{S}(f) \},
\label{eq:overlapvariance}
\end{equation}
where $\widetilde{K} < K$ is an effective number of segments. The left panel of Figure \ref{fig:overlap} depicts the allocation of data points in a sample $x_t$ among segments $x^{(k)}_j$ in Welch's 50\% overlapping segment method.

\subsection{Tapering}
\label{subsec:tapering}

\begin{figure}
    \centering
    \begin{tabular}{ll}
    \includegraphics[width=0.45\textwidth]{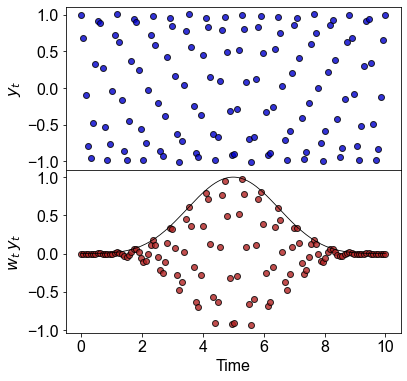} &
    \includegraphics[width=0.54\textwidth]{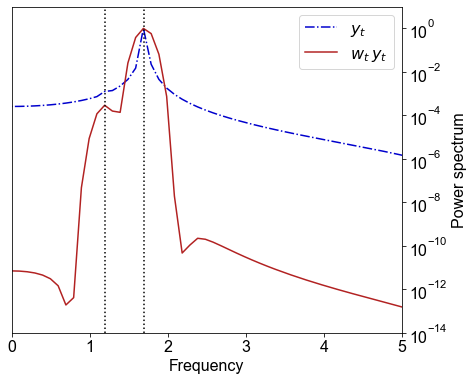} \\
    \end{tabular}
    \caption{{\bf Top left:} Synthetic dataset with $N = 128$, $t_i = 0 \ldots 10$~days, and $y_t = \cos(2 \pi t f_1) + 0.017 \sin(2 \pi t f_2)$, where $f_1 = 1.7$~days$^{-1}$ and $f_2 = 1.2$~days$^{-1}$. {\bf Bottom left:} $w_t y_t$, the product of $y_t$ with a minimum 4-term Blackman-Harris window (gray curve). {\bf Right:} Estimated power spectra of $y_t$ (blue dash-dot line) and $w_t y_t$ (solid red line). Vertical lines mark $f_1$ and $f_2$. Without tapering, leakage from planet 1's strong signal masks the weaker signal from planet 2. When the taper is applied, leakage from planet 1 is confined to frequencies near $f_1$ and the weak signal at $f_2$ is uncovered.}
    \label{fig:taper}
\end{figure}

The value of $\widetilde{K}$ depends on the type of taper (also called a window) applied to each segment. 
Tapers are functions $w_t$ that are pre-multiplied with $x_t$ and $y_t$ to minimize spectral leakage, and are especially valuable for detecting weak signals in the neighborhood of much stronger signals. They are normalized such that $\sum_{t=0}^{N-1} w^2_t = 1$ so as to conserve power. Figure \ref{fig:taper} shows a synthetic RV dataset from a star with two unequal-mass planets in circular orbits:
\begin{equation}
    y_t = \cos(2 \pi t f_1) + 0.017 \sin(2 \pi t f_2),
    \label{eq:synthetic_planets}
\end{equation}
where $f_1 = 1.7$~days$^{-1}$ and $f_2 = 1.2$~days$^{-1}$.\footnote{From a dynamical perspective, such a planetary system is unlikely to be stable; we use it here merely for illustrative purposes.} When no taper is applied (i.e.\ the dataset retains its rectangular or ``boxcar'' taper associated with the fact $y_t = 0$ before and after the observing run such that $\hat{S}_{yy}(f) = \frac{1}{N} |\sum_{t = 0}^{N-1} (y_t - \bar{y}) e^{-i2\pi ft}|^2$), one does not detect the signal associated with planet 2 in $\hat{S}_{yy}(f)$. But when a minimum 4-term Blackman-Harris taper \citep[equation 33 of][]{harris78} is applied to $y_t$, so that the power spectrum estimate becomes 
\begin{equation}
\hat{S}^{w}_{yy}(f) = |\sum_{t = 0}^{N-1} w_t (y_t - \bar{y}) e^{-i2\pi ft}|^2,
\label{eq:taperedspectrum}
\end{equation}
planet 2 is detected despite being responsible for only 0.028\% of $\Var y_t \}$. Here $\hat{S}^{w}_{yy}(f)$ is not computed with Welch's algorithm---it is a standard \citet{schuster1898} periodogram. See \S \ref{subsec:yaxis} for reasons why Figure \ref{fig:taper} and all subsequent power spectrum plots have logarithmic y-axes.

Viewing tapering from a frequency domain context, when the time series is evenly spaced with $\Delta t_j = 1$, we have the following \citep[p.\ 186]{percival93}:
\begin{equation}
    \mbox{E} \{\hat{S}^{w}_{yy}(f)\} = \int_{-1/2}^{1/2} \left | \sum_{t=0}^{N-1} w_n e^{-i 2 \pi t (f-g)} \right |^2 S(f) df,
    \label{eq:specconv}
\end{equation}
which is a convolution between the true power spectrum and the \emph{spectral window} $W(f)$, defined as
\begin{equation} \label{eq:specwin}
    W(f) = \left | \sum_{t = 0}^{N-1} w_t e^{- i 2 \pi f t} \right |^2.
\end{equation}
That is, \emph{the bias of the power spectrum estimator comes from convolving the true power spectrum with the spectral window.} Since spectral leakage is created by the ``smearing" effect of the Fourier transform of the spectral window, 
it's best if $W(f)$ resembles a delta function as closely as possible \citep{harris78}. For RV datasets, we calculate $W(f)$ using the adjoint NFFT by replacing $x_j$ with $w_j$ in Equation \ref{eq:NFFT}. Note that when generalizing Equation \ref{eq:specconv} to unevenly spaced time series, one does not strictly obtain a convolution between $S(f)$ and $W(f)$, since in general the NFFT cannot simply be inverted \citep[Appendix D]{scargle82}. However, we will see in \S \ref{sec:application} that peaks in the RV power spectra are shaped like $W(f)$, and will follow \citet{scargle82} in referring $W(f)$ as the ``spectral window.''

\begin{figure}
    \centering
    \includegraphics[width=0.8\textwidth]{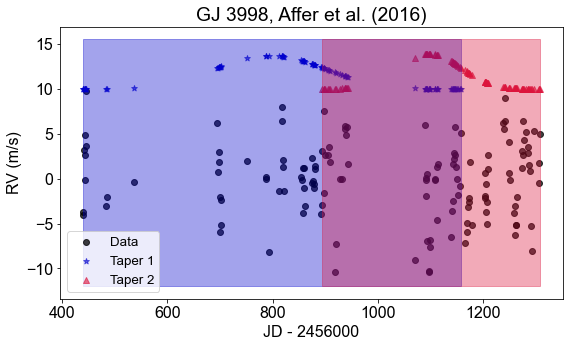}
    \caption{GJ~3998 RV time series (black) divided into two 50\%-overlapping segments shown by blue and orange shading. The dark pink area is where the two segments overlap. The minimum 4-term Blackman-Harris tapers applied to each segment are shown at the top of the plot shifted vertically by +10 for visibility, with blue stars indicating Taper 1 and red triangles indicating Taper 2. RV data come from the HADES survey, which uses the HARPS-N instrument, and were extracted with the TERRA pipeline \citep{affer16}.}
    \label{fig:GJ3998_segments}
\end{figure}

When applying Welch's algorithm, tapering the overlapping segments not only mitigates spectral leakage---it also increases the independence of the various $\hat{S}^{(k)}(f)$. To see why, we examine the right panel of Figure \ref{fig:overlap}, which shows tapers $w_j^{(k)}$ applied to each segment depicted in the left panel of Figure \ref{fig:overlap}. At data point $t = 5$, where $w_j^{(0)}$ is highest,  $w_j^{(1)}$ is nearly zero. Where $w_j^{(1)}$ is at its maximum, overlapping tapers $w_j^{(0)}$ and $w_j^{(2)}$ are near zero, and so on. The tapers ensure that very little information contained in segment $k$ gets repeated in segments $k-1$ or $k+1$. The effective number of segments $\widetilde{K}$ is
\begin{equation}
   \widetilde{K} = \frac{K}{1 + 2 c^2 - 2 c^2 / K},
   \label{eq:ktilde}
\end{equation}
where $c$ is a constant that depends on the type of taper applied \citep{welch67}. The boxcar taper belonging to otherwise untapered segments $x_j^{(k)}$ has $c = 0.5$, while the minimum 4-term Blackman-Harris taper has $c = 0.038$, yielding $\widetilde{K} \approx K$. Figure \ref{fig:GJ3998_segments} shows the GJ~3998 RV data of \citet{affer16} broken into two 50\%-overlapping segments with the associated minimum 4-term Blackman-Harris tapers.

If all RV planet-search data were observed at evenly spaced time intervals, as in the spacecraft example in \S \ref{sec:math}, the benefits of applying tapers to the Welch's segments when computing $\hat{S}_{xx}(f)$, $\hat{S}_{yy}(f)$ and $\hat{C}^2_{xy}(f)$ would far outweigh the drawbacks.\footnote{The high dynamic range delivered by the minimum 4-term Blackman-Harris taper and other similar spectral windows comes at the cost of some resolution in the frequency domain; see \S \ref{subsec:bandwidth} for more on how tapering affects resolution.} Re-examining Figure \ref{fig:taper}, we see that $\hat{S}^{w}_{yy}(f)$---the tapered power spectrum estimate of $y_t$ from Equation \ref{eq:synthetic_planets}---has a dynamic range of over 13 orders of magnitude. This dynamic range would allow planet hunters to identify terrestrial planets, hot Jupiters, rotation, and activity cycles {\it all in the same RV power spectrum estimate} $\hat{S}^{w}_{yy}(f)$: there would be no need to fit the strongest signal, subtract it out, examine a periodogram of the residuals, and keep iterating until no more signals were found. When the temporal cadence is only mildly uneven---as is common in paleoclimatology---the Blackman-Harris window and other similar tapers retain most of their bias-suppression ability \citep[e.g.][]{olafsdottir16}.

\begin{figure}
    \centering
    \includegraphics[width=0.9\textwidth]{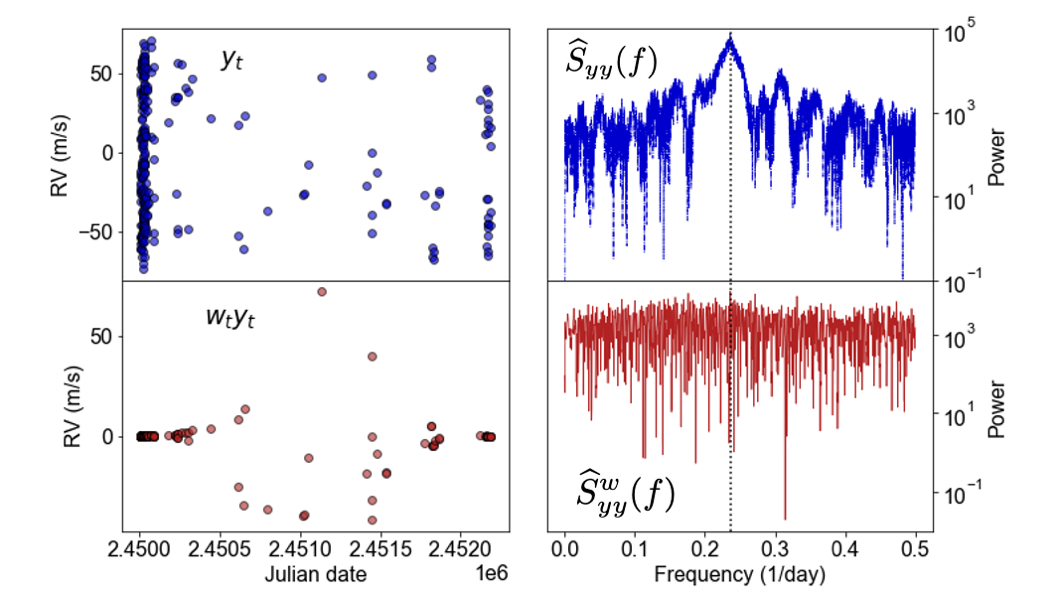}
    \caption{{\bf Top left:} 51~Peg RV time series measured by \citet{butler06}, labeled $y_t$ in our notation. {\bf Bottom left:} $w_t y_t$, the 51~Peg RV time series after applying a minimum 4-term Blackman-Harris taper. {\bf Top right:} $\hat{S}_{yy}(f)$, the estimated power spectrum of $y_t$, shows the strong planet signal at $f = 1 / 4.2308 $~days$^{-1}$ (dotted black line). {\bf Bottom right:} $\hat{S}^{w}_{y}(f)$, the estimated power spectrum of $w_t y_t$, shows only noise.}
    \label{fig:51Pegb}
\end{figure}

But the wildly uneven observing cadence of RV time series often destroys the tapers \citep[e.g.][]{scargle89}. Figure \ref{fig:51Pegb} shows the effect of tapering the 51~Peg~b RV dataset of \citet{butler06}. The top left shows the published data $y_t$, while the bottom left shows $w_t y_t$, where $w_t$ is the minimum 4-term Blackman-Harris taper evaluated at the observation timepoints $t_j$. The top right shows that the planet's signal (black dotted line) is clearly visible in $\hat{S}_{yy}(f)$ (blue dash-dot line), while on the bottom right, $\hat{S}^{w}_{yy}(f)$ shows nothing but noise. Almost all of the 51~Peg observations took place at the very beginning and the very end of $y_t$, when the interpolated $w_t$ is near zero, so tapering removes nearly all the information contained in the time series. When the observing cadence is extremely uneven, we revert to the boxcar tapers when computing $\hat{C}^2_{xy}(f)$. A good rule of thumb is that all tapers should retain the bell-like shapes shown in the right panel of Figure \ref{fig:overlap}. If the ``bell'' is missing huge chunks or its shape is not recognizable when plotted, tapering may do more harm than good.\footnote{Not all useful tapers have bell shapes in the time domain---in particular, higher-order multitapers have zero crossings \citep{s78, t82}---but in this work we use only bell-shaped tapers.}

\subsection{Gaps, weighting, and number of data points per segment}
\label{subsec:gaps}

Another consideration when applying Welch's algorithm to RV data is that large gaps in the middle of a segment should be avoided when possible. Gappy segments are difficult to handle because the resolution limit of $\overline{S}_{xx}(f)$, $\overline{S}_{yy}(f)$, and $\overline{S}_{xy}(f)$ is set by the time duration of the segments:
\begin{equation}
    \mathcal{R}^{(k)} = \frac{1}{t_{N^{(k)}} - t_{0^{(k)}}},
    \label{eq:Rayleigh}
\end{equation}
where $t_{N^{(k)}}$ is the final timestamp in segment $k$, $t_{0^{(k)}}$ is the first timestamp in segment $k$, and $2 \mathcal{R}^{(k)}$---called the Rayleigh resolution in analogy to Rayleigh's criterion in optics---is both the smallest oscillation frequency that can be detected and the frequency separation $\Delta f$ of two barely resolved peaks 
\citep[e.g.][]{godin72}. For evenly spaced data, $\mathcal{R}^{(k)}$ is the same for all segments, 
but RV datasets always yield varying $\mathcal{R}^{(k)}$. A segment $x_j^{(k)}$ that contains one or more large gaps has a low {\it apparent} Rayleigh resolution, which makes one optimistic that low-frequency information missing from other segments might be available in $x_j^{(k)}$. But $\hat{S}_{xx}^{(k)}(f)$ is often misleading at small integer multiples of $\mathcal{R}^{(k)}$ because the gaps make for poor phase coverage of low-frequency oscillations. Here we use the segment with the longest time duration\footnote{The longest-duration segment does not necessarily have the highest number of data points $N^{(k)}$; it is simply the segment with the largest value of $t_{N^{(k)}} - t_{0^{(k)}}$ (Equation \ref{eq:Rayleigh}).} to define the Rayleigh resolution $\mathcal{R}$ of $\overline{S}_{xx}(f)$, $\overline{S}_{yy}(f)$, $\overline{S}_{xy}(f)$, and $\hat{C}^2_{xy}(f)$,
but we urge caution when examining the low-frequency end of each statistic if one or more segments contains a large gap.

In practice it is difficult to keep the Welch's segments gap-free, as seasons and telescope scheduling often combine to isolate a handful of data points from the rest of the time series. The $\alpha$~Cen~B $\log R^{\prime}_{HK}$ time series of \citet{dumusque12} shown in Figure \ref{fig:aCenB_segments} would ideally be broken into four segments, but we use only three segments because the first observing season (modified Julian dates 54550--54650) recorded just 42 observations---not enough to yield a low-bias periodogram \citep{pukkila85, springford2020improving}. No tapers are applied to the $\alpha$~Cen~B data because of the gap in segment 1. With non-overlapping, boxcar-tapered segments, the $\alpha$~Cen~B segmenting scheme differs from Bartlett's method only in that each segment has a different number of data points. For all segmenting patterns, the average cross-spectrum is weighted by the number of data points in each segment:
\begin{equation}
    \bar{S}_{xy}(f) = \frac{\sum_{k=0}^{K-1} N^{(k)} \hat{S}_{xy}^{(k)}(f)}{\sum_{k=0}^{K-1} N^{(k)}}.
    \label{eq:weighting}
\end{equation}
Average periodograms $\bar{S}_{xx}(f)$ and $\bar{S}_{yy}(f)$ are likewise weighted by $N^{(k)}$.

To mitigate the worst manifestations of small-sample bias, we recommend that all segments have $N^{(k)} \geq 100$ \citep{hannan77, pukkila85}, which requires $N \geq 150$ for $K \geq 2$ 50\% overlapping segments. But many published RV datasets have fewer than 100 observations (a practice we do not endorse). Based on the simulations of \citet{das21}, who generated small-sample realizations of AR and ARMA processes and compared the resulting periodograms with the processes' analytically known power spectra, we consider $N^{(k)} = 50$ a hard lower limit. This means the minimum number of astronomical observations required for Welch's algorithm is 75 (two 50\%-overlapping segments each with $N^{(k)} = 50$), though more is much better. (In fact, \citet{th14} present a time series with $N = 1000$ for which the periodogram, which records a power-law turbulent cascade, is still biased by more than seven orders of magnitude.) The transformed coherence $z(f)$
and its noise properties can only be described analytically by Gaussian statistics when $\widetilde{K} \geq 20$ \citep{enochson65, jenkins68}, which requires $N > 1000$. With small $N$, the false positive risk may depart from theoretical expectations in unpredictable ways (see below).



\subsection{Bias correction and false alarm levels}
\label{subsec:bias}

While minimizing bias in $\hat{S}^{(k)}_{xx}(f)$, $\hat{S}^{(k)}_{yy}(f)$, and $\hat{C}^2_{xy}(f)$ is always an important consideration when deploying Welch's algorithm, an approximate bias correction to $\hat{C}^2_{xy}(f)$ is possible. The bias on the magnitude-squared coherence measurement is
\begin{equation}
{\rm bias} \left[ \hat{C}^2_{xy}(f) \right] \approx \frac{\left[ 1 - \hat{C}^2_{xy}(f) \right]^2}{\widetilde{K}}
\label{eq:cohbias}
\end{equation}
\citep{carter73, bendat10}. The debiased coherence estimate is therefore
\begin{equation}
    \hat{C}^{\prime \; 2}_{xy}(f) = \hat{C}^2_{xy}(f) - {\rm bias} \left[ \hat{C}^2_{xy}(f) \right].
    \label{eq:debias}
\end{equation}
All analyses of RV data presented in \S \ref{sec:application} use debiased coherence estimates.

One can also calculate analytical false alarm levels (FALs) for $\hat{C}^{\prime \; 2}_{xy}(f)$:
\begin{equation}
{\rm FAL} = 1 - \alpha^{1/(\widetilde{K}-1)}
\label{eq:falsealarm}
\end{equation}
\citep{carter77, schulz97}, where FAL is the $\hat{C}^{\prime \; 2}_{xy}(f)$ threshold associated with false-alarm probability $\alpha$. Equation \ref{eq:falsealarm} gives false alarm thresholds for $\hat{C}^{\prime \; 2}_{xy}(f)$ given true coherence $C^2_{xy}(f) = 0$, i.e.\ the two time series trace completely unrelated physical phenomena. If a broad-spectrum random process (such as granulation) manifests in both $x_t$ and $y_t$, then Equation \ref{eq:falsealarm} gives artificially low FALs for periodic signals because the two time series have some underlying non-zero coherence at all frequencies. The \texttt{NWelch} software package has an option for calculating frequency-dependent bootstrap FALs for $\hat{C}^{\prime \; 2}_{xy}(f)$ in addition to using Equation \ref{eq:falsealarm}. All FALs presented here are calculated using 10,000 bootstrap iterations.


\subsection{Resolution and the spectral window}
\label{subsec:bandwidth}

While the Rayleigh resolution (Equation \ref{eq:Rayleigh}) is the {\it theoretical} frequency separation between two barely resolved peaks in $\hat{S}_{xx}(f)$, $\hat{S}_{yy}(f)$, or $\hat{C}^{\prime \; 2}_{xy}(f)$, the true frequency resolution of each statistic is determined by the width of the main lobe in the spectral window $W(f)$. Resolution therefore depends on the choice of taper. 
In Figure \ref{fig:taper}, we see that the power spectrum peak associated with planet 1 is quite narrow when the boxcar taper is retained. Although the boxcar taper has poor statistical properties when it comes to leakage and bias, it works well for separating closely spaced signals of similar power. Thus boxcar tapers are useful for asteroseismologists who are trying to resolve modes separated by the small spacing, though they are not optimal for planet hunters who might have to deal with signals of widely varying power. The planet hunter's penalty for using a taper to increase the dynamic range of $\hat{S}_{yy}(f)$ is 
lower resolution in all $\hat{S}(f)$. Following \citet{harris78}, we quantify the half-width of the main lobe in $W(f)$
as the frequency interval $\mathcal{B}$ over which a sinusoidal signal declines from its peak value by 6 deciBels (dB), or a factor of $3.981 (\approx 4)$. One resolution unit is $2 \mathcal{B}$ wide.

\citet{harris78} calculated $\mathcal{B}$ as a function of $\mathcal{R}$ for a variety of tapers applied to evenly spaced time series. For the boxcar taper, $\mathcal{B} = 1.21 \mathcal{R}$.\footnote{Astronomy students may find 1.21 an easy constant to remember, as it's quite similar to the constant in the telescopic Rayleigh resolution equation for a circular aperture, $\theta = 1.22 \lambda / D$ (where $\theta$ is the telescope resolution limit, $\lambda$ is the observation wavelength, and $D$ is the telescope diameter).} The minimum 4-term Blackman-Harris taper has $\mathcal{B} = 2.72 \mathcal{R}$, while the Kaiser-Bessel window (another taper option available in \texttt{NWelch}; see \S \ref{sec:bivariate}) has $\mathcal{B} = 2.39 \mathcal{R}$. The fact that $\mathcal{B}$ is a function of $\mathcal{R}$ means harmonic analysis has a resolution-variance tradeoff in addition to the bias-variance tradeoff discussed in \S \ref{subsec:segmenting}. Users of Welch's method can either prioritize high resolution by constructing a small number of long-duration segments, or emphasize false positive suppression by deploying a larger number of shorter-duration segments. (Of course, small samples and seasonal gaps can constrain the segmenting scheme, making it difficult to find an optimal balance between resolution and variance.) Since almost all RV datasets require segments of varying duration (e.g.\ Figure \ref{fig:aCenB_segments}), we estimate the main lobe half-width of the Welch's estimator as the mean of the main lobe half-widths from the individual segments, weighted by number of points per segment:
\begin{equation}
    \hat{\mathcal{B}} = \frac{\sum_{k = 0}^{K-1} N^{(k)} \mathcal{B}(\mathcal{R})^{(k)}}{\sum_{k=0}^{K-1} N^{(k)}},
    \label{eq:bandwidth}
\end{equation}
where $\mathcal{R}^{(k)}$ is defined in Equation \ref{eq:Rayleigh}.

In practice, the Welch's estimator for unevenly spaced datasets has spectral resolution that depends on the exact timing of the observations, not just the segment durations. (The dependence on observation timing applies to the main-lobe width of the Lomb-Scargle spectral window as well.) It is therefore possible for the actual resolution unit to differ from $2\hat{\mathcal{B}}$ given by Equation \ref{eq:bandwidth}. \texttt{NWelch} allows the user to examine the specific $W(f)$ associated with any Welch's segmenting and tapering scheme.
The software will then empirically calculate $\mathcal{B}$ by finding the frequency at which $W(f) = \mathbf{max} [W(f) / 3.981]$, where $\mathbf{max} [\cdot]$ is the maximum value.

\begin{figure}
    \centering
    \includegraphics[width=0.8\textwidth]{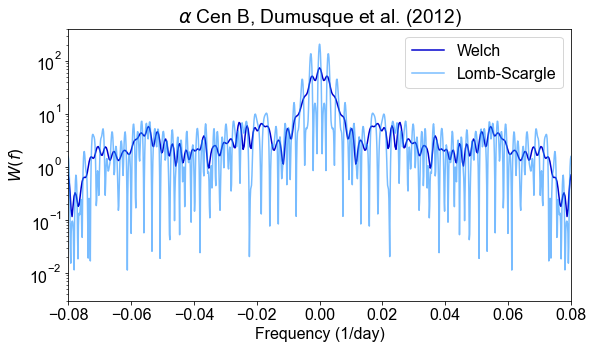}
    \caption{Spectral windows belonging to the \citet{dumusque12} $\alpha$~Cen~B dataset. The ringing in the Lomb-Scargle spectral window (light blue) means each periodic signal yields multiple peaks in $\hat{S}_{xx}(f)$. The Welch's spectral window (dark blue), which has a much cleaner main lobe, will translate each periodic signal into a single power spectrum peak. While this plot zooms in on low frequencies, all spectral windows have non-zero power throughout the frequency domain.}
    \label{fig:specwin}
\end{figure}

We recommend examining $W(f)$ not just to find the resolution of a given Welch's estimator, but because changing the segmenting scheme can strongly alter the shape of the spectral window. For example, Figure \ref{fig:specwin} compares $W(f)$ from the Lomb-Scargle periodogram and the 3-segment Welch's estimator applied to the \citet{dumusque12} $\alpha$~Cen~B dataset (Figure \ref{fig:aCenB_segments}). As we know from \S \ref{subsec:tapering}, a periodic signal does not yield a delta function in $\hat{S}_{xx}(f)$---it instead creates a copy of $W(f)$ centered at the signal frequency. Since the Lomb-Scargle spectral window is beset by ``ringing,'' there can be no clean, isolated peaks in the Lomb-Scargle periodograms of RVs and activity indicators, as noted by \citet[][see also \S \ref{subsec:aCenB}]{rajpaul16}. The Welch's spectral window, which has a well-defined main lobe, is much better suited to identifying periodic signals. Figure \ref{fig:specwin} suggests that spectral window optimization using Welch's method will be a productive avenue for future research.

\subsection{Siegel's test}
\label{subsec:siegel}

\begin{figure}
    \centering
    \begin{tabular}{cc}
    \includegraphics[scale=0.5]{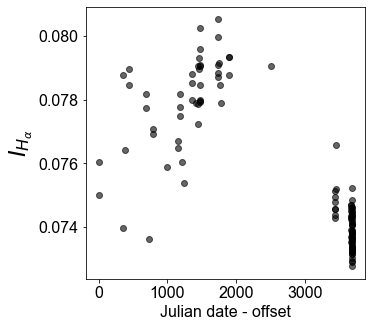} &
    \includegraphics[scale=0.5]{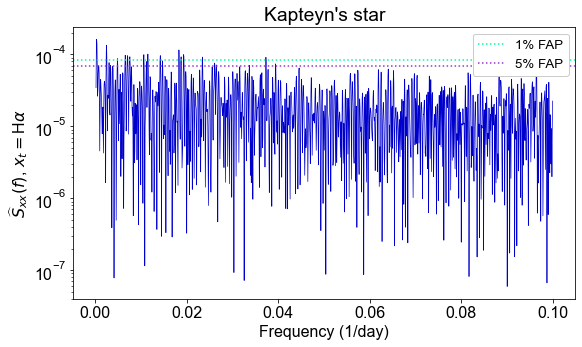} \\
    \multicolumn{2}{c}{ \includegraphics[width=\textwidth]{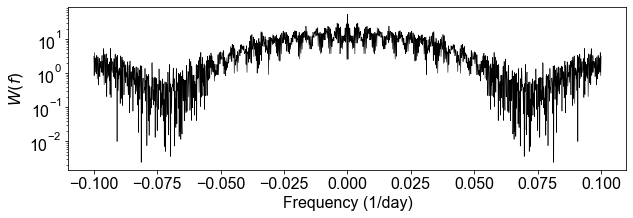} } \\
    \end{tabular}
    \caption{{\bf Top left}: H$\alpha$ indices of Kapteyn's star measured by \citet{robertson15} from the spectra of \citet{angladaescude14}. {\bf Top right:} Lomb-Scargle periodogram of $I_{H_{\alpha}}$ with 5\% and 1\% FALs (purple and green, respectively). The FALs give the unrealistic impression that the periodogram has $\geq 10$ significant peaks. {\bf Bottom:} Spectral window $W(f)$ of the Kapteyn's star $I_{H_{\alpha}}$ time series. The broadness of the main lobe and the large number of spurious spikes suggest that peaks in the Lomb-Scargle periodogram are window function artifacts.}
    \label{fig:kapteyn}
\end{figure}

The last computational method we will describe applies only to power spectrum estimates, not to magnitude-squared coherences, but it is useful for deciding whether a dataset contains periodic signals or just noise. Planet hunters often struggle with unrealistic-looking bootstrap false alarm levels (see \S 2.2 of \citet{cumming04} for information on how to calculate bootstrap FALs). For example, the top left panel of Figure \ref{fig:kapteyn} shows the H$\alpha$-index time series $I_{H_{\alpha}}$ measured by \citet{robertson15} from the Kapteyn's star spectroscopic dataset of \citet{angladaescude14}. The time series yields a Lomb-Scargle periodogram with $\geq 10$ signals that exceed the bootstrap 1\% FAL (Figure \ref{fig:kapteyn}, top right; periodogram computed with \texttt{NWelch}). But should we really believe that a time series with only 112 measurements can record 10 distinct oscillations? It's more likely that the FALS are misleading \citep[indeed, the same false-alarm threshold problem can be seen in the Lomb-Scargle periodogram of the same dataset in Figure 1 of][]{robertson15}, with small-sample statistics, spectral leakage, and aliasing all contributing to the bootstrap failure \citep[e.g.][]{chernick08}. The spectral window shown in the bottom panel of Figure \ref{fig:kapteyn} has a broad main lobe and a large number of spurious spikes, suggesting that $I_{H_{\alpha}}$ periodogram peaks are window function artifacts.

In situations like this, we can deploy Siegel's test for compound periodicity. \citet{siegel80} developed an extension of Fisher's test, which rejects the null hypothesis of white noise when the maximum power in the normalized periodogram exceeds the critical value $g_r$. \citet{percival93} approximate $g_r$ as
\begin{equation}
    g_r \approx 1 - \left( \frac{\alpha}{N_f} \right)^{1 / (N_f-1)},
    \label{eq:fisherg}
\end{equation}
where $N_f$ is the number of entries in the frequency grid and $\alpha$ is the false alarm risk (e.g.\ $\alpha = 0.05$ for 5\% FAP). While \citet{anderson71} notes that Fisher's test is the most powerful identifier of simple periodicity (oscillation at one period only), the test won't work when there are multiple oscillations---such as planet and rotation, rotation and long-term magnetic activity, or rotation with significant power at one or more harmonics. Siegel's test uses a reduced threshold $g = \lambda g_r$, where $\lambda < 1$, and sums all periodogram power in excess of $g$ to compute the test statistic $T_{\lambda}$:
\begin{equation}
    T_{\lambda} = \sum_{j=0}^{N_f-1} \mathbf{max} [ 0, (\hat{S}_{xx}(f_j) - \lambda g_r) ].
    \label{eq:siegelt}
\end{equation}
The value of $T_{\lambda}$ is then compared with a threshold that depends on the number of entries in the frequency grid. If $T_{\lambda}$ exceeds the threshold, the null hypothesis of white noise is rejected and the time series is considered to be periodic. When $\lambda = 0.6$, Siegel's test is conservatively optimized for two periodicities, whereas $\lambda = 0.4$ is sensitive to three or more periodicities but less robust against noise peaks. For the Kapteyn's star periodogram in Figure \ref{fig:kapteyn}, Siegel's test does not reject the null hypothesis of white noise, even with $\lambda = 0.4$: there is no evidence for periodicity in the time series. We will use Siegel's test in \S \ref{sec:application} to see whether the alternative hypothesis of periodicity is supported for certain time series before using those time series to measure rotation periods. Note that Siegel's test doesn't differentiate between a smooth power spectrum with a large dynamic range and a power spectrum consisting of white noise plus a single large oscillation---red noise is a failure mode for both bootstrapping and Siegel's test. We are currently incorporating FALs calculated against a red noise model into \texttt{NWelch} and will discuss these in a future publication.

\section{Application to RV data}
\label{sec:application}

Here we use archival data to demonstrate the use of magnitude-squared coherence in diagnosing stellar RV signals. We select three test datasets that have concurrent RV and activity-indicator time series with $N > 100$. We begin by showing that GJ~581 has significant H$\alpha$-RV coherence at the frequencies of the stellar signals that were misidentified as planets GJ~581~d and GJ~581~g, and use a Welch's power spectrum of the H$\alpha$ index to show that both signals are rotation harmonics. Next we demonstrate that high-frequency stellar signals appear in the magnitude-squared coherences between RV and activity indicators in the $\alpha$~Cen~B dataset assembled by \citet{dumusque12}. We then use coherence between Mt.\ Wilson S-index, H$\alpha$ index, and RV to argue that GJ~3998~b and c may be misdiagnosed stellar activity signals. This section closes with a step-by-step guide to interpreting magnitude-squared coherence measurements.

\subsection{GJ 581}
\label{subsec:GJ581}

GJ~581 became the third M dwarf known to host a planetary system after \citet{bonfils05} used HARPS to discover a Neptune-mass planet with $P = 5.36$~days. \citet{udry07} followed up with reports of planets c ($P = 12.93$~days) and d ($P = 83.6$~days), both in or near the habitable zone. \citet{mayor09} presented another compelling discovery: planet e ($P = 3.15$~days, $M \sin i = 1.9 M_{\oplus}$), one of only a handful of super-Earths discovered in multiplanet systems at the time. They also revised the period of planet d to $66.8$~days, arguing that the one-year alias introduced by Earth's orbit caused some confusion in the \citet{udry07} study. Next, \citet{vogt10} used Keck HIRES data to add planets f ($P = 433$~days) and habitable-zone dweller g ($P = 33.6$~days) to the inventory. The six-planet system with three planets in or near the habitable zone became the object of intense climate modeling efforts \citep[e.g.][]{pierrehumbert11, wordsworth11, vonbloh11, heng11}.

But papers casting doubt on the existence of one or more of the planets began to emerge almost immediately after the report of planets f and g. \citet{angladaescude10} suggested that planet g was an alias of an eccentricity harmonic of planet d, while \citet{gregory11} argued that a Bayesian multiplanet Kepler periodogram could only reliably detect planets b and c. \citet{forveille11} questioned the statistical significance of planets f and g after obtaining 121 new HARPS observations. \citet{baluev13} demonstrated that planets d, f, and g could be artifacts of red noise with a correlation timescale of 10~days. Finally, \citet{robertson14b} demonstrated that ``planet'' d was actually a stellar signal that, when incorrectly modeled in the time domain, created the artifact interpreted as ``planet'' g. Today the Extrasolar Planets Encyclopedia\footnote{exoplanet.eu} states that GJ~581 hosts only three planets: b, c, and e.

\begin{figure}
    \centering
    \includegraphics[width=0.95\textwidth]{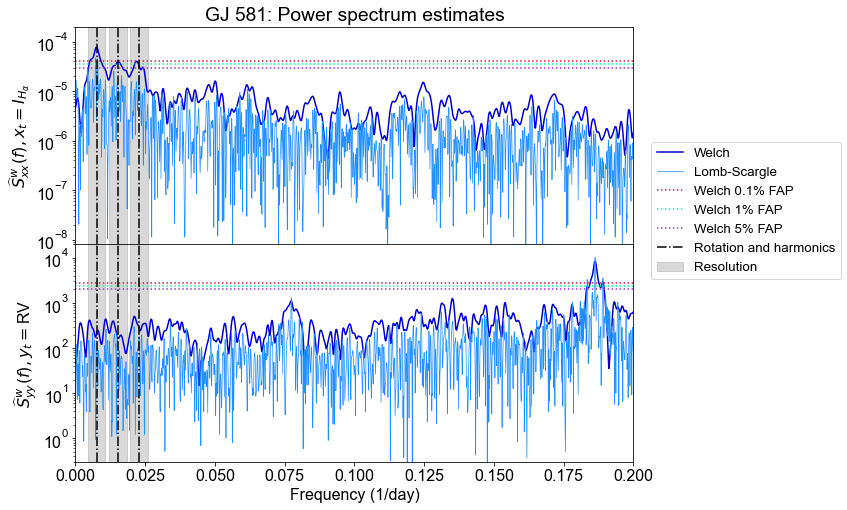}
    \caption{Top: Welch's power spectrum estimate of the GJ~581 $I_{H_{\alpha}}$ time series reported by \citet{robertson14b}, with Lomb-Scargle periodogram plotted for comparison. Bottom: Welch's power spectrum estimate and Lomb-Scargle periodogram of the HARPS GJ~581 RVs reported by \citet{forveille11}. Dotted horizontal lines show 0.1\%, 1\%, and 5\% bootstrap FALs for the Welch's power spectra (Lomb-Scargle FALs are not shown), while vertical black dash-dot lines show the rotation frequency and its first two harmonics. The resolution of the Welch's estimator is indicated by the gray shading.}
    \label{fig:GJ581_periodograms}
\end{figure}

\begin{figure}
    \centering
    \includegraphics[width=0.8\textwidth]{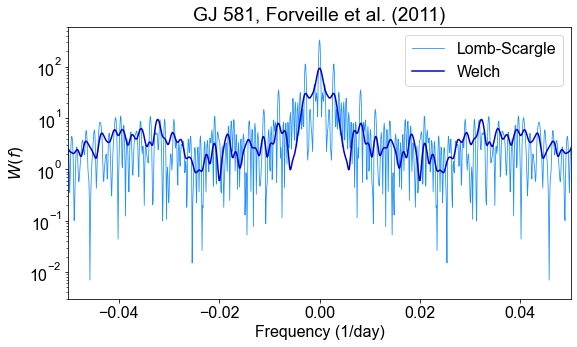}
    \caption{Spectral windows of the generalized Lomb-Scargle periodogram (light blue) and the Welch's power spectrum estimate (dark blue) calculated from the \citet{forveille11} observations of GJ~581. The Welch's power spectrum estimate was created using three 50\%-overlapping segments with minimum 4-term Blackman-Harris tapers.}
    \label{fig:GJ581_spectral_window}
\end{figure}

Since \citet{robertson14b} identified signals d and g as rotation artifacts, we will start by measuring the star's rotation period. Figure \ref{fig:GJ581_periodograms} shows Welch's power spectrum estimates of the time series used in the \citet{robertson14b} analysis: $\hat{S}^w_{xx}(f)$ given $x_t = I_{H_{\alpha}}$ (top) and $\hat{S}^w_{yy}(f)$ given $y_t =$RV (bottom). Both power spectrum estimates were computed using three 50\% overlapping segments with the minimum 4-term Blackman-Harris taper applied to each segment. As with the $\alpha$~Cen~B dataset, the Welch's estimates have a much cleaner spectral window than the generalized Lomb-Scargle periodograms (Figure \ref{fig:GJ581_spectral_window}). \citet{robertson14b} found a primary peak in the $I_{H_{\alpha}}$ Lomb-Scargle periodogram at $P_1 = 125$~days plus a secondary peak at $P_2 = 138$~days and attributed the split peak to phase changes in the rotation signal. However, the signals at $f_1 = 1/P_1$ and $f_2 = 1/P_2$ are not quite separated by $2 \mathcal{R}$ in the $I_{H_{\alpha}}$ generalized Lomb-Scargle periodogram, so cannot be said to be truly distinct. The Welch's power spectrum estimate shows a strong single peak that exceeds the 0.1\% FAL at $P = 132$~days ($f = 0.00758$~days$^{-1}$), which we take to be the true rotation period. There are also significant peaks at the first two rotation harmonics, $2 f_{\rm rot}$ and $3 f_{\rm rot}$. Although spectral window of the Welch's estimator is wide, the rotation signal and its harmonics are each separated by more than one resolution unit. The Welch's RV power spectrum has a statistically significant peak at the orbital frequency of planet b ($f = 0.186$~days$^{-1}$) plus a local maximum at the frequency of planet c ($f = 0.0774$~days$^{-1}$), but no obvious signals at the rotation frequency or its harmonics. The conservative Siegel's test with $\lambda = 0.6$ finds a $\geq 95$\% chance that both time series are periodic, suggesting that the bootstrap FALs are realistic.

\begin{figure}
    \centering
    \includegraphics[width=0.8\textwidth]{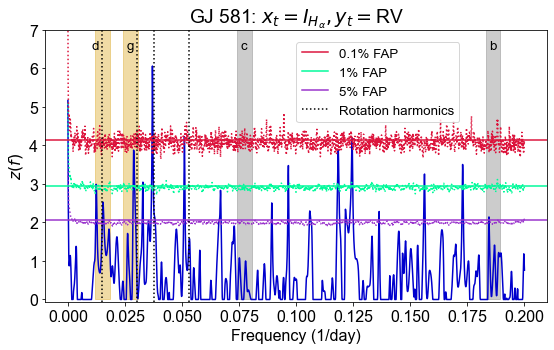}
    \caption{Magnitude-squared coherence estimate from HARPS spectra, with RV measured by \citet{forveille11} and H$\alpha$ index measured by \citet{robertson14b}. Solid horizontal lines show false alarm levels from Equation \ref{eq:falsealarm}, while dotted lines in the same color scheme show bootstrap false alarm levels. The stellar signals originally identified as planets d and g are shown by yellow shaded regions of width $2\mathcal{B}$. Gray shaded regions of width $2\mathcal{B}$ surround the orbital frequencies of planets b and c. Vertical dotted black lines show rotation harmonics at which $z(f)$ exceeds the 1\% FAL.}
    \label{fig:GJ581_coherence}
\end{figure}

\citet{robertson14b} used several lines of reasoning to argue that ``planets'' d and g were really stellar signals: they created a separate fit to RV as a function of $I_{H_{\alpha}}$ for each observing season, identified correlations between $I_{H_{\alpha}}$ and bisector inverse slope, and calculated a new RV model after subtracting off the best-fit seasonal straight-line models RV($I_{H_{\alpha}}$). But a single calculation of $\hat{C}^{\prime ; 2}_{xy}(f)$ is enough to reveal the stellar origins of the two signals. Figure \ref{fig:GJ581_coherence} shows transformed magnitude-squared coherence $z(f)$ given $x_t = I_{H_{\alpha}}$, $y_t =$RV. 
Gray bands of width $2\mathcal{B}$ are centered at the orbital frequencies of planets b and c. The yellow bands, also of width $2\mathcal{B}$, are centered at the frequencies of ``planets'' d and g reported by \citet{vogt10}. Vertical black dotted lines show rotation harmonics $f = n f_{\rm rot}$ (where $n$ is an integer) near which $z(f)$ exceeds the 5\% FAL. ``Planet'' d sits right atop the first rotation harmonic ($f = 2 f_{\rm rot}$) at the center of a resolution unit that includes two statistically significant coherence peaks, one of which exceeds the 1\% FAL. ``Planet'' g is within half a resolution unit of the third rotation harmonic ($f = 4 f_{\rm rot}$), which is also the location of a coherence signal that almost reaches the 0.1\% FAL. Significant $I_{H_{\alpha}}$-RV coherence can also be seen at other rotation harmonics: at both $f = 5 f_{\rm rot}$ and $f = 7 f_{\rm rot}$, $z(f)$ exceeds the 0.1\% FAL. In contrast, the band centered on planet c is clean. The band centered on planet b includes a signal that rises above the 5\% FAL. That signal is likely to be a false positive---indeed, we expect to see more than one false positive above the 5\% FAL given that our coherence estimate has a frequency range far greater than 20 resolution units---but further study of the GJ~581 system that incorporates other activity indicators might be useful.

The strong signals at the rotation harmonics in Figures \ref{fig:GJ581_periodograms} and \ref{fig:GJ581_coherence} suggest that GJ~581 has a complex rotation signal. Spectral power and coherence at high-order harmonics may result from the star hosting multiple large spots or spot groups \citep[e.g.][]{rodono86, gunther20, perger21, perugini21}. In the next section, we will investigate coherent stellar signals that do not appear to be associated with the dominant rotation period, but which may be artifacts of differential rotation, giant cells, or supergranulation.

\subsection{$\alpha$~Cen~B}
\label{subsec:aCenB}

The $\alpha$~Cen~B dataset assembled by \citet{dumusque12} has proven to be useful for identifying spectroscopic signatures of stellar activity. While the star is generally quiet \citep{cincunegui07}, it developed one or more large spot groups that caused obvious rotational modulation in the $\log R^{\prime}_{HK}$ time series from 2010 March 23 to 2010 June 12 \citep{thompson17}. This modulation is visible in Segment 2 of Figure \ref{fig:aCenB_segments}. \citet{wise18} used the 2010 March-June data to find activity-sensitive absorption lines with modulation in either half-depth range or core flux, while \citet{thompson17} used the same data to identify pseudo-emission features with rotationally driven RV changes. Here we will search for stellar signals using Welch's power spectra of the activity indicators and magnitude-squared coherence between activity indicators and RV.

\begin{figure}
    \centering
    \includegraphics[width=\textwidth]{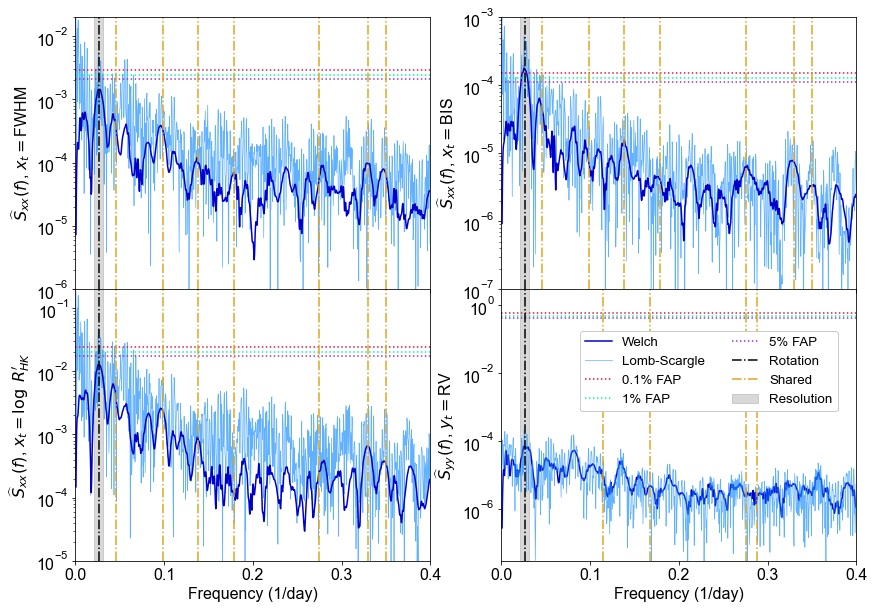}
    \caption{Welch's power spectra of the \citet{dumusque12} $\alpha$~Cen~B dataset: FWHM (top left), BIS (top right), $\log R^{\prime}_{HK}$ (bottom left), and RV (bottom right). Lomb-Scargle periodograms are shown in light blue for comparison. Dotted horizontal lines show 0.1\%, 1\%, and 5\% bootstrap false alarm thresholds for the Welch's power spectrum (Lomb-Scargle false alarm thresholds are not shown). The black dash-dot line denotes the star rotation period, while the gray shaded region shows the resolution limit of the Welch's power spectrum estimate. Vertical dash-dot yellow lines show shared oscillations identified via magnitude-squared coherence.}
    \label{fig:aCenB_periodograms}
\end{figure}

Figure \ref{fig:aCenB_periodograms} shows Welch's power spectra of full width at half maximum of the cross correlation between the spectrum and a digital mask \citep[FWHM;][top left]{pepe00}, bisector velocity span \citep[BIS;][top right]{toner88}, $\log R^{\prime}_{HK}$ (bottom left), and RV (bottom right) calculated using the segmenting scheme in Figure \ref{fig:aCenB_segments}. Before applying Welch's algorithm to the RV data, we removed the binary motion by subtracting the best-fit quadratic model \citep{endl16}. To suppress spectral leakage from the long-period activity cycle, each $x^{(k)}_j$ of FWHM, BIS, and $\log R^{\prime}_{HK}$ and each $y^{(k)}_j$ of RV
had a linear trend removed before $\hat{S}_{xx}^{(k)}(f)$ was calculated. The combination of segmenting and detrending mimics the low-pass filtering applied by \citet{dumusque12} and the Gaussian process model constructed by \citet{suarezmascareno17b}. 
Figure \ref{fig:aCenB_periodograms} also shows Lomb-Scargle periodograms in light blue. The Welch and Lomb-Scargle periodograms differ at low $f$ partly because the Welch's estimator contains no information about signals with periods longer than the longest segment duration, and partly because of the detrending.\footnote{See \S \ref{sec:bivariate} for a ``like-for-like'' comparison between the single-segment, untapered \texttt{NWelch} power spectrum and the \texttt{astropy.timeseries.LombScargle} periodogram.}

Although the Welch's power spectrum of BIS is the only one that has a signal that exceeds the bootstrap white-noise 5\% false alarm threshold, Siegel's test finds that Welch's power spectrum estimates of all of the activity-indicator time series (FWHM, BIS, $\log R^{\prime}_{HK}$) show periodicity at the 95\% significance level (\S \ref{subsec:siegel}). 
We therefore average the frequencies of maximum power in the BIS, FWHM, and $\log R^{\prime}_{HK}$ periodograms to find the rotation period that best describes the \citet{dumusque12} dataset: $P_{\rm rot} = 37.7$~days ($f_{\rm rot} = 0.0265$~day$^{-1}$), which is consistent with rotation period estimates from the literature \citep{dewarf10, brandenburg17, suarezmascareno17b}.\footnote{Since rotation signals are quasiperiodic, it's common for estimates of the rotation period to vary slightly with activity cycle phase and/or number of spot groups on the star surface \citep[e.g.][]{gilbert21}, hence the variety of similar but not identical measurements from the literature.} We will use our measured rotation period as a benchmark when searching for stellar signals. Vertical black dotted lines in Figure \ref{fig:aCenB_periodograms} show shared oscillations identified in the magnitude-squared coherences (see below).

\begin{figure}
    \centering
    \includegraphics[width=0.9\textwidth]{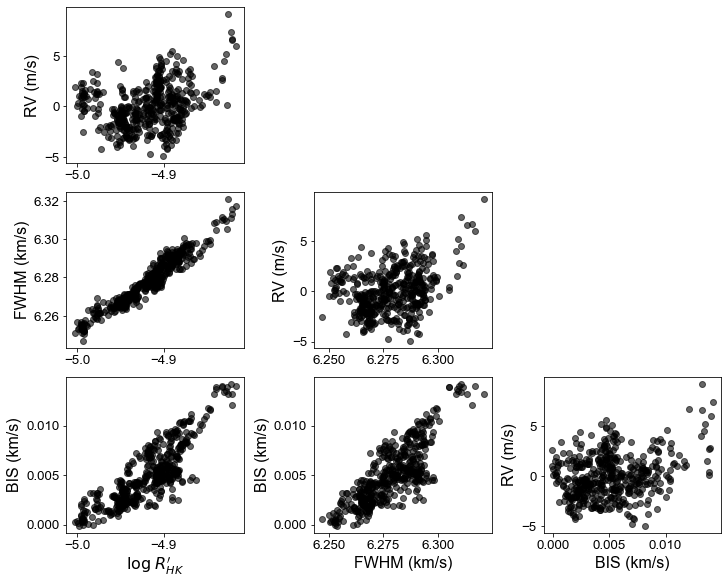}
    \caption{Scatter plots $y_t$ vs.\ $x_t$ for all four $\alpha$~Cen B observables. While there is no obvious straight-line relationship between RV and any activity indicator, the indicators are tightly correlated amongst themselves, suggesting that they trace the same physical processes.}
    \label{fig:corner}
\end{figure}

Before we examine any magnitude-squared coherence estimates, we pause to consider Figure \ref{fig:corner}, which shows scatter plots $y_t$ vs.\ $x_t$ for all ${4 \choose 2} = 6$ pairs of observables. RV is not well described as a straight-line function of any activity indicator. We can optimistically hope that's because activity signals are not manifesting in RV, but it's possible that (a) the relationships between RV and activity indicators are nonlinear, phase-lagged (Figure \ref{fig:example1}), and/or noisy, or (b) $\log R^{\prime}_{HK}$, FWHM, and BIS do not provide a complete description of stellar activity. 
On the other hand, the activity indicators show obvious straight-line relationships with {\it each other}---particularly $\log R^{\prime}_{HK}$ and FWHM---which suggests that they all trace the same underlying physical processes. In a magnitude-squared coherence analysis with two activity indicators as $x_t$ and $y_t$, we should expect to see statistically significant values in $\hat{C}^{\prime \; 2}_{xy}(f)$ and $z(f)$.

\begin{figure}
    \centering
    \includegraphics[width=0.8\textwidth]{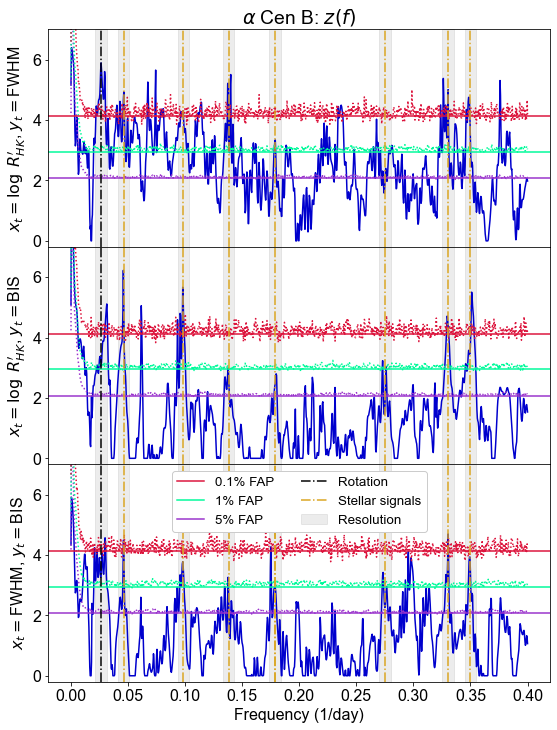}
    \caption{Hyperbolic arctangent-ransformed activity indicator coherences from the \citet{dumusque12} $\alpha$~Cen~B data. Top: $z(f)$ given $x_t = \log R^{\prime}_{HK}, y_t =$FWHM; middle: $z(f)$ given $x_t = \log R^{\prime}_{HK}, y_t =$BIS; bottom: $z(f)$ given $x_t =$FWHM, $y_t =$BIS. The black dash-dotted line shows the shared rotation signal, while the gray shaded area represents the resolution limit $\mathcal{B}$. Yellow dash-dot lines show signals that exceed the 1\% false alarm threshold in at least two of the three panels.}
    \label{fig:aCenB_indicator_coherences}
\end{figure}

In Figure \ref{fig:aCenB_indicator_coherences}, we see our prediction of high coherences between the activity indicators borne out. The plots show $z(f)$, the $\operatorname{atanh}$-transformed magnitude-squared coherence, given $x_t = \log R^{\prime}_{HK}$, $y_t =$FWHM (top); $x_t = \log R^{\prime}_{HK}$, $y_t =$BIS (middle); and $x_t =$FWHM, $y_t =$BIS (bottom). Solid horizontal lines show false alarm thresholds calculated with Equation \ref{eq:falsealarm}, while dotted horizontal lines show bootstrap false alarm thresholds (\S \ref{sec:bivariate}). The rotation frequency is marked with a black dash-dot line, with the resolution limit indicated in gray. As expected, all three plots show statistically significant $z(f)$ across the entire rotation band. Other frequencies besides rotation at which $z(f)$ exceeds the 1\% false alarm threshold in at least {\it two of three} panels in Figure \ref{fig:aCenB_indicator_coherences} are marked with vertical yellow dash-dot lines; if a signal at frequency $f$ is significant in two of three coherences between activity indicators, it means the oscillation is present in all three indicators. Coherent stellar signals are located at $f = (0.0460, 0.0985, 0.138, 0.179, 0.275, 0.330, 0.350)$~days$^{-1}$, or $P = (21.7, 10.2, 7.27, 5.60, 3.64, 3.03, 2.86)$~days.

We know from Figure \ref{fig:specwin} that the shared signals are not window function artifacts, but their physical origin is unclear: most of them do not fall at simple rotation harmonics ($f = n f_{\rm rot}$, where $n$ is an integer), though $f = 0.138$~days$^{-1}$ is within half a resolution unit of the $n = 4$ harmonic. Differential rotation sometimes creates secondary peaks in periodograms of photometry and activity indicators \citep{reinhold13}; our best guess is that we are seeing a secondary peak, perhaps along with some beating between differential rotation signals. Supergranulation and giant cells may be in play on the shorter timescales \citep{nordlund09}.
If we re-examine the Welch's power spectrum estimates of $\log R^{\prime}_{HK}$, FWHM, and BIS in Figure \ref{fig:aCenB_periodograms}, we see that the shared signals (again marked by yellow dash-dot lines) are all at or near local maxima. Even though none of those local maxima appear statistically significant when compared with our bootstrap white-noise false alarm thresholds, the underlying oscillations may be real. The downward slopes of $\hat{S}_{xx}(f)$ for $x_t =$FWHM, BIS, and $\log R^{\prime}_{HK}$ suggest that false alarm thresholds in the $\alpha$~Cen~B power spectra should be computed from a red noise model \citep[e.g.][]{olafsdottir16}. We are developing this functionality and will include it in a future release of \texttt{NWelch}. Red noise has been incorporated in models of RV data by (e.g.) \citet{baluev13}, \citet{tuomi13}, and \citet{feng16}.

\begin{figure}
    \centering
    \includegraphics[width=0.8\textwidth]{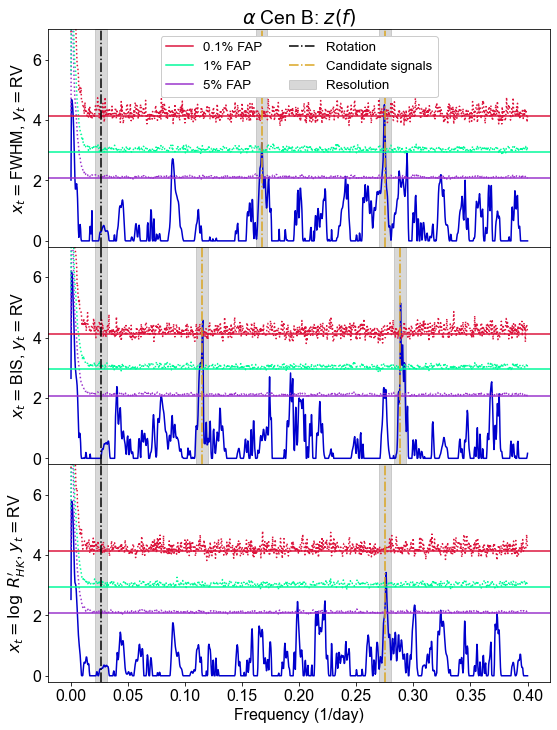}
    \caption{Transformed magnitude-squared coherence for $x_t =$FWHM, $y_t =$RV (top); $x_t = BIS$, $y_t =$RV (middle); and $x_t = \log R^{\prime}_{HK}$, $y_t =$RV (bottom). The color scheme follows Figure \ref{fig:aCenB_indicator_coherences}. Yellow dash-dot lines mark shared oscillations that rise above the 1\% FAL.}
    \label{fig:aCenB_RV_coherences}
\end{figure}

Now we turn to the magnitude-squared coherences between activity indicators and RV. Figure \ref{fig:aCenB_RV_coherences} shows $z(f)$ given $y_t =$RV and $x_t =$FWHM (top), $y_t =$RV and $x_t =$BIS (middle), and $y_t =$RV and $x_t = \log R^{\prime}_{HK}$ (bottom). The color scheme is the same as in Figure \ref{fig:aCenB_indicator_coherences}. Magnitude-squared coherences in the rotation band are near zero, indicating that no activity indicator is a good tracer of the way rotation manifests throughout the {\it entire} duration of the RV dataset. 
In each panel of Figure \ref{fig:aCenB_RV_coherences}, signals that exceed the 1\% FAL in the $z(f)$ estimate shown in that panel {\it only} are marked with yellow dash-dot lines. The activity indicators' shared signal at $f = 0.275$~days$^{-1}$ (Figure \ref{fig:aCenB_indicator_coherences}) also shows up in coherences between FWHM \& RV and $\log R^{\prime}_{HK}$ \& RV. FWHM-RV coherence has a second peak at $f = 0.167$~days$^{-1}$ ($P = 5.99$~days, top panel). BIS-RV coherence has peaks at $f = 0.115$~days$^{-1}$ and $f = 0.288$~days$^{-1}$ (P = $8.70$~days and $P = 3.47$~days, middle panel). The four coherent RV-activity indicator oscillations are marked in the Welch's RV power spectrum in Figure \ref{fig:aCenB_periodograms} (lower-right panel). The FWHM-RV oscillation at $f = 0.167$~days$^{-1}$ coincides with a local maximum of the Welch's RV power spectrum. The $\log R^{\prime}_{HK}$ power spectrum also has a local maximum at $f = 0.167$~days$^{-1}$, though not the FWHM power spectrum. The coherent RV-BIS oscillation at $f = 0.115$~days$^{-1}$ is within one resolution unit of peaks in the FWHM and RV power spectra. All four power spectra have a local maximum at $f = 0.275$~days$^{-1}$.

Our analysis of the \citet{dumusque12} $\alpha$~Cen~B dataset shows that periodic short-timescale stellar activity occurs at other frequencies besides $f_{\rm rot}$ and its harmonics. We have identified seven oscillatory signals that are present in all three activity indicators, plus four coherent activity indicator-RV oscillations. Figure \ref{fig:corner} demonstrates that the stellar RV signals could not have been diagnosed by fitting a straight-line model to RV as a function of an activity indicator.
Accurately identifying the activity signals is an important step toward suppressing false positives in RV planet searches.

\subsection{GJ 3998}
\label{subsec:GJ3998}

Since 2012, the HArps-N red Dwarf Exoplanet Survey (HADES) program has been surveying 78 early-type M dwarfs for Doppler shifts induced by rocky planets \citep{perger17a}. The first HADES discovery paper featured two planets orbiting GJ~3998 with periods $P_b = 2.65$~days and $P_c = 13.7$~days \citep{affer16}. Aware of the tendency of M dwarfs to be more active than solar-type couterparts of the same age, the HADES team examined time series of $I_{H_{\alpha}}$ and Mt.\ Wilson S-index measured from the same spectra as the RVs, as well as near-simultaneous EXORAP and APACHE photometry. \citet{affer16} identified two signals with periods $P_1 = 30.7$~days and $P_2 = 42.5$~days that were present in RV, $I_{H_{\alpha}}$, and S-index and posited that $P_1$ was the true rotation period while $P_2$ represented modulation due to differential rotation. The photometry was consistent with rotation period $P_1$. Following \citet{robertson14b}, the HADES team verified that the planetary signals were present in generalized Lomb-Scargle periodograms of RV data from all observing seasons and demonstrated that the Spearman’s rank correlation coefficients of RV as a function of S-index and $I_{H_{\alpha}}$ were insignificant. They then concluded that the signals at $P_b$ and $P_c$ were planetary in origin.

\begin{figure}
    \centering
    \includegraphics[width=0.95\textwidth]{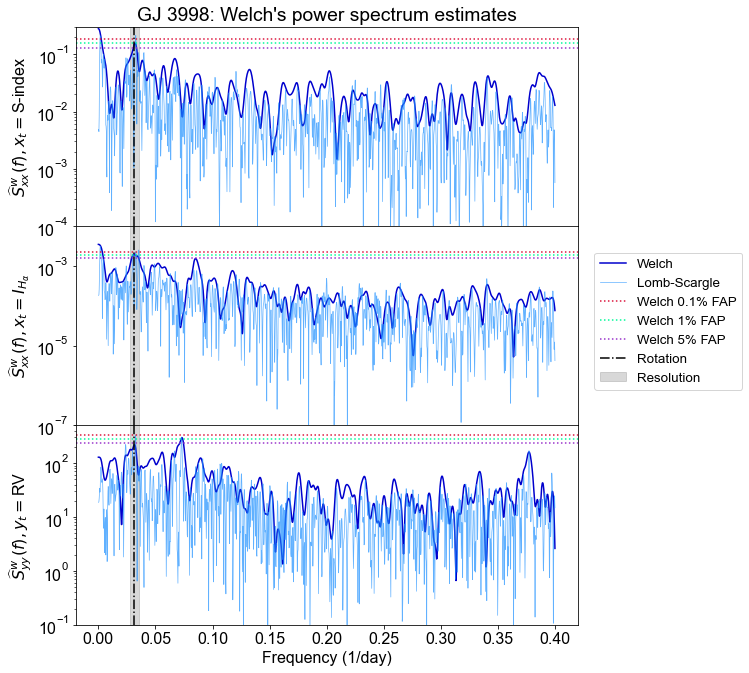}
    \caption{Welch's (dark blue) and generalized Lomb-Scargle (light blue) periodograms of the GJ~3998 spectroscopic data of \citet{affer16}. The color scheme is the same as in Figure \ref{fig:GJ581_periodograms}.}
    \label{fig:GJ3998_periodograms}
\end{figure}

As with GJ~581 and $\alpha$~Cen~B, we begin our analysis by measuring the star rotation period from Welch's power spectrum estimates with the segmentation and tapering scheme shown in Figure \ref{fig:GJ3998_segments}. \citet{affer16} explain how the traditional HARPS and HARPS-N way of measuring RVs, by cross-correlating with a mask consisting of a thin rectangle surrounding each spectral line center, is suboptimal for M dwarfs because their spectra feature substantial line blending. Accordingly, in Figure \ref{fig:GJ3998_periodograms} we examine the Welch's and generalized Lomb-Scargle periodograms from the RV, S-index, and $I_{H_{\alpha}}$ time series measured with the TERRA pipeline \citep{angladaescude12}. The S-index and $I_{H_{\alpha}}$ time series contain statistically significant peaks at 0.0321~days$^{-1}$ and 0.0311~days$^{-1}$, respectively, which we average together to find $f_{\rm rot} = 0.0316$~days$^{-1}$ / $P_{\rm rot} = 31.7$~days. Our rotation period is similar to $P_1$ identified by \citet{affer16}. Although the stellar signal with frequency $f_2 = 1/P_2$ would be more than one resolution unit away from $f_{\rm rot}$, the Welch's power spectra do not show any evidence for such a signal.

\begin{figure}
    \centering
    \includegraphics[width=0.98\textwidth]{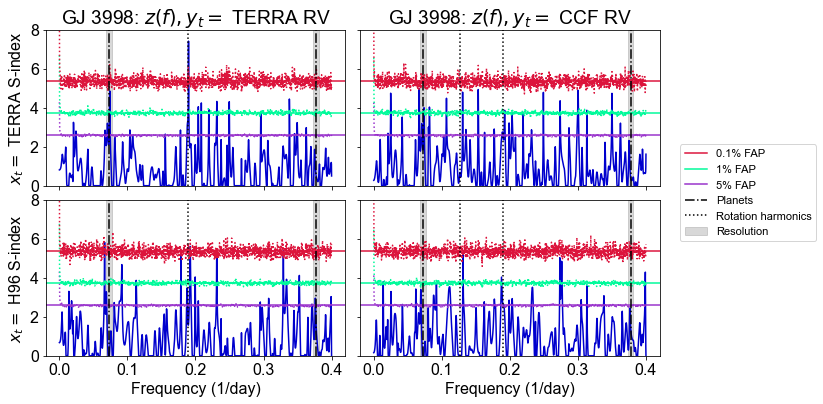}
    \caption{Transformed magnitude-squared coherences between S-index and RV from the HADES GJ~3998 observations of \citet{affer16}. Top left: $x_t = $TERRA S-index, $y_t = $TERRA RV. Top right: $x_t = $TERRA S-index, $y_t = $CCF RV. Bottom left: $x_t = $H96 S-index, $y_t = $TERRA RV. Bottom right: $x_t = $H96 S-index, $y_t = $CCF RV. All panels show a signal at the frequency of planet c. In addition, all $z(f)$ estimates show high coherence at the fifth rotation harmonic ($f = 6 f_{\rm rot}$), while $z(f)$ estimates with $y_t = $CCF RV have high coherence at the third harmonic ($f = 4 f_{\rm rot}$). Rotation harmonics are marked by dotted black lines.} 
    \label{fig:GJ3998_S_coherence}
\end{figure}

We now turn to magnitude-squared coherence. Although \citet{affer16} emphasize that the TERRA pipeline is preferred over the CCF-based HARPS DRS pipeline for M dwarfs, they nevertheless present RVs measured with both methods. They also present a second set of activity indicator time series, with S-index measured according to \citet[H96]{henry96} and $I_{H_{\alpha}}$ measured as in \citet[R13]{robertson13}. Accordingly, we analyze coherences with $x_t =$~TERRA S-index, H96 S-index, TERRA $I_{H_{\alpha}}$, and R13 $I_{H_{\alpha}}$, and with $y_t =$~TERRA RV and CCF RV---eight measurements of $z(f)$ in all. Figure \ref{fig:GJ3998_S_coherence} shows the four transformed coherence measurements with an S-index measurement as $x_t$. Planets b and c are marked by vertical dash-dot black lines surrounded by a shaded region indicating the resolution unit. In three of four S-RV coherences, there is a peak within half a resolution unit of planet c that exceeds the 1\% FAL. The fourth, with $x_t = $H96 S-index and $y_t = $CCF RV, has a signal at $f_c$ that exceeds the 5\% FAL. As with GJ~581, we are also seeing high-order rotation harmonics: all $z(f)$ estimates have high coherence at the fifth harmonic ($f = 6 f_{\rm rot}$), while $z(f)$ estimates with $y_t = $CCF RV have high coherence at the third harmonic ($f = 4 f_{\rm rot}$).

\begin{figure}
    \centering
    \includegraphics[width=0.98\textwidth]{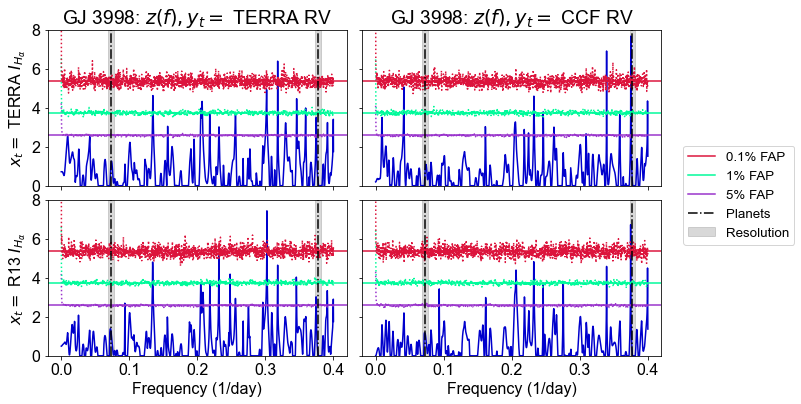}
    \caption{Transformed magnitude-squared coherences between $I_{H_{\alpha}}$ and RV from the HADES GJ~3998 observations of \citet{affer16}. Top left: $x_t = $TERRA $I_{H_{\alpha}}$, $y_t = $TERRA RV. Top right: $x_t = $TERRA $I_{H_{\alpha}}$, $y_t = $CCF RV. Bottom left: $x_t = $R13 $I_{H_{\alpha}}$, $y_t = $TERRA RV. Bottom right: $x_t = $R13 $I_{H_{\alpha}}$, $y_t = $CCF RV. All panels show a stellar signal at the frequency of planet b, with FAP$\ll 1$\% for $y_t = $CCF RV.}
    \label{fig:GJ3998_Ha_coherence}
\end{figure}

If we examine the coherences with an $I_{H_{\alpha}}$ measurement as $x_t$ plotted in Figure \ref{fig:GJ3998_Ha_coherence}, we find evidence for a stellar signal at the frequency of planet b. When $y_t = $CCF RV, the $I_{H_{\alpha}}$-RV coherence at $f_b$ exceeds the 0.1\% FAL. With the TERRA RVs, we find $I_{H_{\alpha}}$-RV coherence over the 5\% FAL at $f_b$ and nearing the 1\% FAL for $x_t = $TERRA $I_{H_{\alpha}}$. However, the band surrounding planet c is clean. $I_{H_{\alpha}}$ does not have any coherent oscillations with RV directly at the rotation harmonics, though there is a coherence peak almost within a half resolution unit of $4 f_{\rm rot}$ for $y_t = $TERRA RV. The reason \citet{affer16} could not see a straight-line relationship between either $I_{H_{\alpha}}$ and RV or S-index and RV is because there is more than one oscillation shared between RV and the activity indicators---in addition to the planet candidates, rotation harmonics are present, and so are higher-frequency signals that are not obviously associated with rotation, as in the $\alpha$~Cen~B data.

While we are not yet ready to state definitively that the GJ~3998 RVs show only stellar signals, the system requires further follow-up in light of our results. If coherence between RV and activity indicators at $f_b$ and $f_c$ persists after additional data are taken, that would suggest that the planets are not real. It's also crucial to measure multiple activity indicators, since they may not all trace the same underlying physical phenomena. Indeed, \citet{robertson13} highlight the fact that Ca H\&K and H$\alpha$ emission come from different chromospheric depths and point out that signal-to-noise ratio is often problematic in the Ca H\&K lines in M dwarf spectra.

\subsection{Interpreting magnitude-squared coherence measurements}

What should observers look for when applying bivariate frequency-domain techniques to their own planet-search datasets? In our analyses of GJ~581, $\alpha$~Cen~B, and GJ~3998, we roughly followed the following procedure:

\begin{enumerate}

\item Examine the spectral window of the Welch's estimator. Look for sidelobes that could yield false positives in the power spectra. Calculate the Rayleigh resolution limit, the analytical resolution unit $\mathcal{B}$ \citep[e.g.][]{harris78}, and the empirical value of $\mathcal{B}$ for a dataset's specific observing cadence. \texttt{NWelch} automatically reports these numbers for each segmenting/tapering scheme. Be aware that two signals with a frequency separation of less than $2 \mathcal{B}$ are not statistically distinguishable. It's especially important to consider resolution when searching for planets near the star rotation period or differential rotation.

\item Use Welch's power spectra of the activity indicators to estimate the rotation period. While many planet-search targets will have previous rotation period measurements in the literature, the quasiperiodic nature of the rotation signal means one might recover a somewhat different rotation period depending on the observational epoch and activity cycle phase \citep{robertson14b}. It's important to determine how rotation is manifesting in the particular dataset under analysis. For the $\alpha$~Cen~B dataset, in which three activity indicators yielded three slightly different rotation period estimates, we averaged together the various estimates (\S \ref{subsec:aCenB}). It's possible that the community will come up with a more sophisticated way to handle differing rotation period measurements.

\item Search for coherence between RV and activity indicators within one resolution unit of the rotation period, the periods of planet candidates, and the harmonics of all of the above. For any planet that is not large enough or close enough to the star to trigger stellar activity, $\hat{C}^{\prime \; 2}_{xy}(f)$ and $z(f)$ should be impervious to eccentricity harmonics. Observers who find coherent RV-activity indicator signals at either their planet candidate frequency or its harmonics should use extreme caution before reporting a planet discovery.

It's unclear how high up the overtone sequence we should expect to see rotation harmonics---for example, if there's a peak in $z(f)$ near the eighth harmonic ($9 f_{\rm rot}$), is it truly rotation-related, or is its proximity to a harmonic just a coincidence? We hope stellar physicists will explore the complexity of rotation signals and the extent to which their harmonics should be traceable by RV datasets. We also hope stellar physicists will weigh in on possible sources of coherent RV-activity signals that are not related to rotation.

\item When coherent RV-activity indicator signals are found, check to see whether they line up with local maxima in the power spectra. Sometimes a power spectrum peak that's not significant when judged against a white noise model indicates a real signal. The signal processing literature features more sophisticated ways of checking the significance of power spectrum peaks, such as F-testing \citep[e.g.][]{thomson94}, false alarm thresholds determined from red noise models \citep[e.g.][]{olafsdottir16}, and prewhitening followed by checking against the exponential quantiles. But these techniques are mostly unexplored for RV data. When a peak in $z(f)$ corresponds to a local maximum in the RV and/or activity-indicator power spectrum, it lends support to the hypothesis that the coherence is real and not spurious. However, given that the geophysics literature features coherent signals that don't correspond to power spectrum local maxima \citep[see example in][]{pardoiguzquiza12}, one shouldn't be too quick to dismiss any signals in $z(f)$ as false positives.

\end{enumerate}

\section{Conclusions and Plans for Future Work}
\label{sec:conclusions}

Magnitude-squared coherence is a powerful tool for diagnosing stellar signals in bivariate RV-activity indicator time series. By combining Welch's power spectrum estimates with magnitude-squared coherence measurements, we were able to identify the signals that were originally labeled GJ~581~d, g as rotation harmonics. In the $\alpha$~Cen~B dataset, we mapped bivariate oscillations onto non rotation-related local maxima in the Welch's power spectrum estimates of FWHM, BIS, $\log R^{\prime}_{HK}$, and RV. Finally, in the GJ~3998 data, we found high coherence between $I_{H_{\alpha}}$, S-index, and RV at the frequencies of planet candidates b and c. Since it is now standard practice for planet hunters to analyze and publish activity indicator time series along with RVs \citep[e.g.][]{dalba21, gonzalezalvarez21, maldonado21}, and since most stellar signals show up in some, but not all, activity indicators (\S \ref{subsec:GJ3998}), we recommend that every planet discovery be vetted by analyzing the magnitude-squared coherence between RV and as many activity indicators as can be measured. Doing so is computationally cheap, and our \texttt{NWelch} software package is publicly available from a repository that contains examples of all functionality (\S \ref{sec:bivariate}). More generally, since the nonuniform Fourier transform can map to any set of frequencies, the frequency domain approach described here is the \emph{only} way to properly describe correlations between lagged versions of two time series with unequal observing cadence---there is no time-domain approach to the problem that does not involve interpolation.

The Welch's estimator that underlies our coherence measurements gives cleaner spectral windows and lower variance than generalized Lomb-Scargle periodograms (\S \ref{subsec:GJ581}), even when tapers are not applied to the segments (\S \ref{subsec:aCenB}). Magnitude-squared coherence is valuable primarily for its ability to identify stellar signals, but sometimes it is the spectral window and not the star that is responsible for false positives \citep[e.g.][]{rajpaul16}. Our results suggest that Welch's method may be a valuable addition to the set of frequency-domain methods already in use in planet searches, such as the generalized Lomb-Scargle periodogram \citep[GLS,][]{zechmeister09}, the Bayesian GLS \citep{mortier15}, and the maximum-likelihood periodogram \citep{stoica89}. Again, it is computationally inexpensive to compute a Welch's power spectrum estimate with \texttt{NWelch}, and we recommend that all planet hunters closely examine the Welch's power spectra that are generated along with magnitude-squared coherence estimates.

In \S \ref{subsec:tapering} we demonstrate how tapering can increase the dynamic range of the power spectrum estimator. The ability to see low-amplitude and high-amplitude signals in the same power spectrum estimate, without iteratively fitting and subtracting out signals one by one, would be invaluable in searches for earthlike planets. However, in both statistics and astronomy, almost all research on tapering has been confined to datasets with even or near-even observing cadence \citep[e.g.][]{chave2019multitaper, springford2020improving}. Quantifying the dynamic range attainable by the Welch's power spectrum estimator for different uneven observing cadences, taper types (e.g.\ Blackman-Harris, Kaiser-Bessel, Hann), and segmenting schemes would be a useful direction for future research.

Our findings show that star rotation signals can be complex, with significant RV-activity indicator coherence appearing at high-order rotation harmonics in our two example M~dwarfs (\S \S \ref{subsec:GJ581}, \ref{subsec:GJ3998}). Furthermore, the $\alpha$~Cen~B dataset has multivariate oscillations with unclear physical origins. At the short end of the period range ($P \sim 3$~days), the multivariate oscillations with may be associated with supergranulation or giant cells. The longer-period oscillations ($P = 7.27, 10.2, 21.7$~days) could be related to differential rotation and harmonics thereof---though the period separation between our measured rotation period of 37.7~days and the 21.7-day signal is above the $\sim 30$\% threshold selected by \citet{reinhold13} for detecting differential rotation in {\it Kepler} targets. We expect investigations of magnitude-squared coherence between RV and activity indicators to yield new insights into stellar physics.

The false positive rate of the magnitude-squared coherence estimator applied to RV data should be explored further. We have not yet encountered an RV dataset without statistically significant coherence between RV and an activity indicator at {\it some} frequency. This is to be expected: the precision achieved by the RV community in the past decade makes it inevitable that stellar signals will manifest in RV data instead of being subsumed by instrument noise, and even if that weren't true, coherences have spurious signals just like any other statistic. But we do not yet know the extent to which broad-spectrum processes such as turbulence boost the RV-activity indicator coherence and complicate the interpretation of analytical false alarm thresholds in $\hat{C}^{\prime \; 2}_{xy}(f)$ and $z(f)$. Furthermore, RV datasets have an observing cadence that is far more uneven than the paleoclimatology datasets on which the nonuniform Welch's method is typically deployed, so astronomers might be dealing with different false positive rates than geophysicists. We expect the RV community to achieve better understanding of false positive rates as the magnitude-squared coherence estimator is applied to more datasets.





\acknowledgments

We are grateful to Lily Zhao, Debra Fischer, and the EXPRES team for allowing us to test our statistical methods on new EXPRES data. We thank Joan Caicedo Vivas and Henry Sanford-Crane for asking important questions about our methodology and Catherine Lembo and Rebecca Hutchinson for input on early phases of this work. Effort by SDR, VRD, and JH was funded by Bartol Research Institute. VRD received additional funding from the UNIDEL foundation. This material is based upon work supported by the U.S. Department of Energy, Office of Science, under contract number DE-AC02-06CH11357.

%

\vspace{5mm}
\facilities{Keck (HIRES), ESO La Silla 3.6m (HARPS), Telescopio Nazionale Galileo (HARPS-N)}


\software{NWelch \citep{dodsonrobinson22}, \\
          FINUFFT \citep{barnett19, barnett21}, \\
          astropy \citep{astropy}, \\ 
          redfit-x \citep{olafsdottir16}
          }
          
\appendix

\section{\texttt{NWelch} software package}
\label{sec:bivariate}

All Welch's power spectrum and coherence estimates in this work were computed with the new \texttt{NWelch} software package \citep{dodsonrobinson22}, written for \texttt{python} 3, which performs Fourier analysis of bivariate time series. \texttt{NWelch} borrows heavily from \texttt{redfit-x}, a package for cross-spectral analysis of paleoclimate time series written in \texttt{fortran} \citep{olafsdottir16}. \texttt{NWelch} is stored in a repository at \texttt{https://github.com/sdrastro/NWelch}; see the README for installation instructions and a complete listing of the software functionality. The software consists of two classes: a base class called \texttt{TimeSeries}, which implements the univariate calculations, and a derived class called \texttt{Bivariate}, which calculates $\hat{C}^{\prime \; 2}_{xy}(f)$, $z(f)$, and $\hat{\phi}(f)$. \texttt{Jupyter} notebooks containing the calculations shown in \S \ref{sec:application} are located in the repository directories \texttt{GJ581}, \texttt{aCenB}, and \texttt{GJ3998}. The \texttt{demo} directory contains an example \texttt{Jupyter} notebook that demonstrates the full functionality of the \texttt{TimeSeries} class using the Barnard's star activity indicator dataset of \citet{toledopadron19}. 

\subsection{Power spectrum plots: logarithmic or linear y-axis?}
\label{subsec:yaxis}

\texttt{NWelch} defaults to a logarithmic y-axis for all power spectrum plots. The user can select a linear y-axis (see the demo notebooks in the \texttt{github} repository for instructions), but there are good reasons to use logarithmic scaling \citep[e.g.][]{thomson94}:
\begin{itemize}
\item Power spectrum peaks that aren't significant when judged against a white noise model can correspond to coherence peaks that {\it are} significant, so it's good to be able to see all the local maxima in the power spectrum (see the $\alpha$ Cen B analysis in \S \ref{subsec:aCenB}),
\item Human vision and hearing respond to the intensity of stimuli on a logarithmic scale,
\item A downward slope in $\log_{10}[\hat{S}_{xx}(f)]$, which is easy to spot on a semilog-y plot, is an indicator of red noise,
\item Multiplicative effects appear additive on a logarithmic scale, so convolution in the time domain---which is equivalent to multiplication in the frequency domain---yields simple addition in $\log_{10}[\hat{S}_{xx}(f)]$.
\end{itemize}
Finally, the convention in the signal processing literature is to use semilog-y plots for power spectra because engineering processes often use linear filters such as moving averages or Gaussian smoothing. A moving average applied to time-domain data is obvious on a semilog-y plot of the power spectrum because the plot will show the filter sidelobes. We recommend that all astronomers working in the frequency domain examine semilog-y plots of their power spectra.


\subsection{Nonuniform Fourier transforms \label{app:NUFFT}}

The workhorse calculation that underlies all \texttt{NWelch} functionality is the non-uniform fast Fourier transform (NFFT). Here we provide a brief overview of the mathematics behind the NFFT.
\citet{lomb76} and \citet{scargle82} were the first to develop periodogram (Fourier
spectrum) analysis techniques that were quite powerful for finding and testing the significance of weak periodic
signals in otherwise random, unevenly sampled data.
Given a set of data values $x_j= 1,\ldots,N$ at respective observation times, the
Lomb-Scargle periodogram is constructed as follows. First, compute the data’s mean and variance
by
\begin{equation}
  \bar{x} = \frac{1}{N} \sum_j x_j; \quad \hat{\sigma}^2_x = \frac{1}{N-1} \sum_j (x_j - \bar{x})^2 
\end{equation}
 Second, for each angular frequency $\omega = 2 \pi f >0$ of interest, compute a time-offset $\tau$ by
\begin{equation}
  \tan 2 \omega \tau = \frac{ \sum_j \sin 2 \omega t_j}{\sum_j \cos 2 \omega t_j}
\end{equation}
 Third, the Lomb-Scargle normalized periodogram (spectral power as a function of angular frequency $\omega = 2 \pi f$) is
 defined by
 \begin{equation} \label{eq:LS}
   P_N(\omega) = \frac{1}{2\sigma^2} 
                 \left[ \frac{ \left[\sum_j (x_j-\bar{x}) \cos \omega (t_j - \tau)\right]^2}{\sum_j
(x_j-\bar{x}) \cos^2 \omega (t_j - \tau)} 
                 + \frac{ \left[\sum_j (x_j-\bar{x}) \sin \omega (t_j - \tau)\right]^2}{\sum_j
(x_j-\bar{x}) \sin^2 \omega (t_j - \tau)} \right]
\end{equation}
The constant $\tau$ makes $P_N(\omega)$ completely independent of shifting all the $t_j$ by
any constant. Lomb (1976) showed that this particular choice of offset has another,
deeper, effect: it makes equation \ref{eq:LS} identical to the equation that one
would obtain if one estimated the harmonic content of a data set, at a given
frequency $\omega$ by linear least-squares fitting to the model
 \begin{equation*}
   x(t) - \bar{x} = A \cos \omega t + B \sin \omega t.
 \end{equation*}
 
Lomb-Scargle periodograms can be approximated by employing an ``extirpolation"
process (interpolation of the unevenly spaced points onto a regular grid) using the
Lagrange interpolation method, and then using the ordinary FFT \citep{press1989fast}.
However, as described in \citet{leroy2012fast} \citep[\S 7]{springford2020improving},
one can employ the adjoint nonuniform FFT \citep{keiner2009using} for a fast
implementation which is not approximate. Briefly, one computes 
\begin{align}
S_y &= \sum_i (x_i - \bar{x}) \sin \omega t_i, 
\quad & C_y &= \sum_i (x_i - \bar{x}) \cos \omega t_i, \\
S_2 & = \sum_i \sin 2 \omega t_i, 
\quad & C_2  &= \sum_i \cos 2 \omega t_i,
\end{align}
then
\begin{align}
\sum_i (x_i - \bar{x}) \cos\omega(t_i - \tau) &= C_y \cos \omega\tau + S_y \sin \omega\tau, \\
\sum_i (x_i - \bar{x}) \sin\omega(t_i - \tau) &= S_y \cos \omega\tau - C_y \sin \omega\tau, \\
\sum_i \cos 2  \omega(t_i - \tau) & = \frac{N}{2} + \frac{1}{2} C_2 \cos 2\omega\tau + \frac{1}{2}
S_2 \sin 2\omega\tau,\\
\sum_i \sin 2 \omega(t_i - \tau) &= \frac{N}{2}  - \frac{1}{2} C_2 \cos 2\omega\tau - \frac{1}{2}
S_2 \sin 2\omega\tau.
\end{align}
\texttt{NWelch} uses FINUFFT, the Flatiron Institute non-uniform fast Fourier transform,\footnote{https://finufft.readthedocs.io/en/latest/index.html} to compute the adjoint NFFT. Methods for performing the exponential sums are given by \citet{barnett19} and \citet{barnett20}.
For all power spectrum estimates (Figures \ref{fig:GJ581_periodograms}, \ref{fig:aCenB_periodograms} and \ref{fig:GJ3998_periodograms}), \texttt{NWelch} uses the power spectral density normalization:
\begin{equation}
    \Var x_t \} = \sum_{m=0}^{N_f-1} \hat{S}_{xx}(f_m) \Delta f,
\end{equation}
where $m$ is the integer index of the frequency grid and $\Delta f$ is the frequency grid spacing. 

\subsection{Comparison with \texttt{astropy.timeseries.LombScargle}}

\begin{figure}
    \centering
    \includegraphics[width=0.8\textwidth]{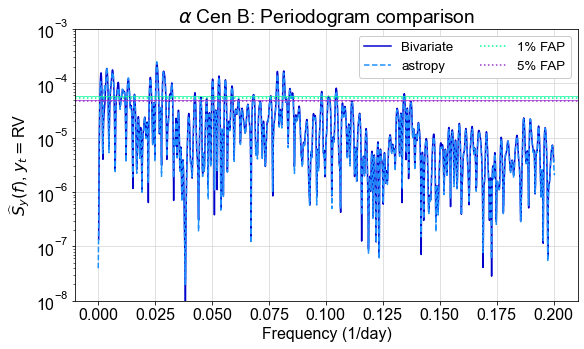}
    \caption{\texttt{NWelch} single-segment power spectrum estimate (dark blue, solid line) and \texttt{astropy} Lomb-Scargle periodogram (light blue, dashed line) of the $\alpha$~Cen~B RVs measured by \citet{dumusque12}. 5\% and 1\% false alarm thresholds are shown in purple and green, respectively, with the solid horizontal lines showing \texttt{astropy} false alarm thresholds and the dotted horizontal lines showing \texttt{NWelch} false alarm thresholds. The two power spectrum estimates and their bootstrap false alarm thresholds are nearly identical.}
    \label{fig:percompare}
\end{figure}

The most basic \texttt{NWelch} task is to compute a periodogram of an unevenly spaced time series without tapering, detrending, or segmentation. Here we show that this task produces results that are consistent with \texttt{astropy.timeseries.LombScargle}. The top panel of Figure \ref{fig:percompare} shows two periodograms of the residuals of the \citet{dumusque12} $\alpha$~Cen~B RV data after subtracting a quadratic model of binary motion. The periodogram plotted with a dark blue, solid line comes from \texttt{NWelch}, while the periodogram plotted with the light blue, dashed line comes from \texttt{astropy.timeseries.LombScargle}. The periodograms are almost identical, as are their bootstrap 5\% and 1\% false alarm levels (purple and green, respectively, with dotted lines for \texttt{NWelch} FALs and solid lines for \texttt{astropy.timeseries.LombScargle} FALs). For simple Lomb-Scargle periodograms with bootstrap false alarm thresholds, \texttt{NWelch} and \texttt{astropy.timeseries.LombScargle} can be used interchangeably. However, \texttt{NWelch} does not include the \citet{baluev08} method for calculating false alarm probabilities based on extreme value theory.

\bibliographystyle{aasjournal}
\typeout{}
\bibliography{biblio}

\begin{thebibliography}{}
\expandafter\ifx\csname natexlab\endcsname\relax\def\natexlab#1{#1}\fi
\providecommand{\url}[1]{\href{#1}{#1}}
\providecommand{\dodoi}[1]{doi:~\href{http://doi.org/#1}{\nolinkurl{#1}}}
\providecommand{\doeprint}[1]{\href{http://ascl.net/#1}{\nolinkurl{http://ascl.net/#1}}}
\providecommand{\doarXiv}[1]{\href{https://arxiv.org/abs/#1}{\nolinkurl{https://arxiv.org/abs/#1}}}

\bibitem[{{Affer} {et~al.}(2016){Affer}, {Micela}, {Damasso}, {Perger},
  {Ribas}, {Su{\'a}rez Mascare{\~n}o}, {Gonz{\'a}lez Hern{\'a}ndez}, {Rebolo},
  {Poretti}, {Maldonado}, {Leto}, {Pagano}, {Scandariato}, {Zanmar Sanchez},
  {Sozzetti}, {Bonomo}, {Malavolta}, {Morales}, {Rosich}, {Bignamini},
  {Gratton}, {Velasco}, {Cenadelli}, {Claudi}, {Cosentino}, {Desidera},
  {Giacobbe}, {Herrero}, {Lafarga}, {Lanza}, {Molinari}, \& {Piotto}}]{affer16}
{Affer}, L., {Micela}, G., {Damasso}, M., {et~al.} 2016, \aap, 593, A117,
  \dodoi{10.1051/0004-6361/201628690}

\bibitem[{Anderson(1971)}]{anderson71}
Anderson, T.~W. 1971, The Statistical Analysis of Time Series (John Wiley \&
  Sons)

\bibitem[{{Anglada-Escud{\'e}} \& {Butler}(2012)}]{angladaescude12}
{Anglada-Escud{\'e}}, G., \& {Butler}, R.~P. 2012, \apjs, 200, 15,
  \dodoi{10.1088/0067-0049/200/2/15}

\bibitem[{{Anglada-Escud{\'e}} \& {Dawson}(2010)}]{angladaescude10}
{Anglada-Escud{\'e}}, G., \& {Dawson}, R.~I. 2010, arXiv e-prints,
  arXiv:1011.0186.
\newblock \doarXiv{1011.0186}

\bibitem[{{Anglada-Escude} {et~al.}(2014){Anglada-Escude}, {Arriagada},
  {Tuomi}, {Zechmeister}, {Jenkins}, {Ofir}, {Dreizler}, {Gerlach}, {Marvin},
  {Reiners}, {Jeffers}, {Butler}, {Vogt}, {Amado}, {Rodriguez-Lopez},
  {Berdinas}, {Morin}, {Crane}, {Shectman}, {Thompson}, {Diaz}, {Rivera},
  {Sarmiento}, \& {Jones}}]{angladaescude14}
{Anglada-Escude}, G., {Arriagada}, P., {Tuomi}, M., {et~al.} 2014, \mnras, 443,
  L89, \dodoi{10.1093/mnrasl/slu076}

\bibitem[{{Astropy Collaboration} {et~al.}(2013){Astropy Collaboration},
  {Robitaille}, {Tollerud}, {Greenfield}, {Droettboom}, {Bray}, {Aldcroft},
  {Davis}, {Ginsburg}, {Price-Whelan}, {Kerzendorf}, {Conley}, {Crighton},
  {Barbary}, {Muna}, {Ferguson}, {Grollier}, {Parikh}, {Nair}, {Unther},
  {Deil}, {Woillez}, {Conseil}, {Kramer}, {Turner}, {Singer}, {Fox}, {Weaver},
  {Zabalza}, {Edwards}, {Azalee Bostroem}, {Burke}, {Casey}, {Crawford},
  {Dencheva}, {Ely}, {Jenness}, {Labrie}, {Lim}, {Pierfederici}, {Pontzen},
  {Ptak}, {Refsdal}, {Servillat}, \& {Streicher}}]{astropy}
{Astropy Collaboration}, {Robitaille}, T.~P., {Tollerud}, E.~J., {et~al.} 2013,
  \aap, 558, A33, \dodoi{10.1051/0004-6361/201322068}

\bibitem[{{Baluev}(2008)}]{baluev08}
{Baluev}, R.~V. 2008, \mnras, 385, 1279,
  \dodoi{10.1111/j.1365-2966.2008.12689.x}

\bibitem[{{Baluev}(2013)}]{baluev13}
---. 2013, \mnras, 429, 2052, \dodoi{10.1093/mnras/sts476}

\bibitem[{Barnett(2020)}]{barnett20}
Barnett, A.~H. 2020, Aliasing error of the exp$(\beta \sqrt{1-z^2})$ kernel in
  the nonuniform fast Fourier transform.
\newblock \doarXiv{2001.09405}

\bibitem[{Barnett(2021)}]{barnett21}
---. 2021, Applied and Computational Harmonic Analysis, 51, 1,
  \dodoi{https://doi.org/10.1016/j.acha.2020.10.002}

\bibitem[{Barnett {et~al.}(2019)Barnett, Magland, \& af~Klinteberg}]{barnett19}
Barnett, A.~H., Magland, J., \& af~Klinteberg, L. 2019, SIAM Journal on
  Scientific Computing, 41, C479, \dodoi{10.1137/18M120885X}

\bibitem[{{Bartlett}(1948)}]{bartlett48}
{Bartlett}, M.~S. 1948, \nat, 161, 686, \dodoi{10.1038/161686a0}

\bibitem[{Bendat \& Piersol(2010)}]{bendat10}
Bendat, J.~S., \& Piersol, A.~G. 2010, Random Data: {A}nalysis and Measurement
  Procedures (Wiley)

\bibitem[{{Boisse} {et~al.}(2011){Boisse}, {Bouchy}, {H{\'e}brard}, {Bonfils},
  {Santos}, \& {Vauclair}}]{boisse11}
{Boisse}, I., {Bouchy}, F., {H{\'e}brard}, G., {et~al.} 2011, \aap, 528, A4,
  \dodoi{10.1051/0004-6361/201014354}

\bibitem[{{Bonfils} {et~al.}(2005){Bonfils}, {Forveille}, {Delfosse}, {Udry},
  {Mayor}, {Perrier}, {Bouchy}, {Pepe}, {Queloz}, \& {Bertaux}}]{bonfils05}
{Bonfils}, X., {Forveille}, T., {Delfosse}, X., {et~al.} 2005, \aap, 443, L15,
  \dodoi{10.1051/0004-6361:200500193}

\bibitem[{Bonfils {et~al.}(2007)Bonfils, Mayor, Delfosse, Forveille, Gillon,
  Perrier, Udry, Bouchy, Lovis, Pepe, {et~al.}}]{bonfils07}
Bonfils, X., Mayor, M., Delfosse, X., {et~al.} 2007, Astronomy \& Astrophysics,
  474, 293, \dodoi{doi:10.1051/0004-6361:20077068}

\bibitem[{{Bortle} {et~al.}(2021){Bortle}, {Fausey}, {Ji}, {Dodson-Robinson},
  {Ramirez Delgado}, \& {Gizis}}]{bortle21}
{Bortle}, A., {Fausey}, H., {Ji}, J., {et~al.} 2021, \aj, 161, 230,
  \dodoi{10.3847/1538-3881/abec89}

\bibitem[{{Bourrier} {et~al.}(2018){Bourrier}, {Dumusque}, {Dorn}, {Henry},
  {Astudillo-Defru}, {Rey}, {Benneke}, {H{\'e}brard}, {Lovis}, {Demory},
  {Moutou}, \& {Ehrenreich}}]{bourrier18}
{Bourrier}, V., {Dumusque}, X., {Dorn}, C., {et~al.} 2018, \aap, 619, A1,
  \dodoi{10.1051/0004-6361/201833154}

\bibitem[{{Brandenburg} {et~al.}(2017){Brandenburg}, {Mathur}, \&
  {Metcalfe}}]{brandenburg17}
{Brandenburg}, A., {Mathur}, S., \& {Metcalfe}, T.~S. 2017, \apj, 845, 79,
  \dodoi{10.3847/1538-4357/aa7cfa}

\bibitem[{Bronez(1992)}]{bronez92}
Bronez, T.~P. 1992, ieeesp, 40, 2941

\bibitem[{{Butler} {et~al.}(2006){Butler}, {Wright}, {Marcy}, {Fischer},
  {Vogt}, {Tinney}, {Jones}, {Carter}, {Johnson}, {McCarthy}, \&
  {Penny}}]{butler06}
{Butler}, R.~P., {Wright}, J.~T., {Marcy}, G.~W., {et~al.} 2006, \apj, 646,
  505, \dodoi{10.1086/504701}

\bibitem[{{Butler} {et~al.}(2017){Butler}, {Vogt}, {Laughlin}, {Burt},
  {Rivera}, {Tuomi}, {Teske}, {Arriagada}, {Diaz}, {Holden}, \&
  {Keiser}}]{butler17}
{Butler}, R.~P., {Vogt}, S.~S., {Laughlin}, G., {et~al.} 2017, \aj, 153, 208,
  \dodoi{10.3847/1538-3881/aa66ca}

\bibitem[{Carter(1977)}]{carter77}
Carter, G. 1977, IEEE Transactions on Acoustics, Speech, and Signal Processing,
  25, 90

\bibitem[{Carter(1987)}]{carter87}
Carter, G.~C. 1987, procieee, 75, 236

\bibitem[{Carter {et~al.}(1973)Carter, Knapp, \& Nuttall}]{carter73}
Carter, G.~C., Knapp, C.~H., \& Nuttall, A. 1973, IEEE Transactions on
  Acoustics, Speech, and Signal Processing, 21, 337

\bibitem[{Chave(2019)}]{chave2019multitaper}
Chave, A.~D. 2019, Geophysical Journal International, 218, 2165

\bibitem[{{Chave} {et~al.}(1992){Chave}, {Luther}, {Lanzerotti}, \&
  {Medford}}]{chave92}
{Chave}, A.~D., {Luther}, D.~S., {Lanzerotti}, L.~J., \& {Medford}, L.~V. 1992,
  \grl, 19, 1411, \dodoi{10.1029/92GL01286}

\bibitem[{Chernick(2008)}]{chernick08}
Chernick, M.~R. 2008, Bootstrap methods: A guide for practitioners and
  researchers, 2nd edn. (Wiley-Interscience)

\bibitem[{{Cincunegui} {et~al.}(2007){Cincunegui}, {D{\'\i}az}, \&
  {Mauas}}]{cincunegui07}
{Cincunegui}, C., {D{\'\i}az}, R.~F., \& {Mauas}, P.~J.~D. 2007, \aap, 469,
  309, \dodoi{10.1051/0004-6361:20066503}

\bibitem[{{Cumming}(2004)}]{cumming04}
{Cumming}, A. 2004, \mnras, 354, 1165, \dodoi{10.1111/j.1365-2966.2004.08275.x}

\bibitem[{{Dalba} {et~al.}(2021){Dalba}, {Kane}, {Li}, {MacDougall},
  {Rosenthal}, {Cherubim}, {Isaacson}, {Thorngren}, {Fulton}, {Howard},
  {Petigura}, {Schwieterman}, {Peluso}, {Esposito}, {Marchis}, \&
  {Payne}}]{dalba21}
{Dalba}, P.~A., {Kane}, S.~R., {Li}, Z., {et~al.} 2021, \aj, 162, 154,
  \dodoi{10.3847/1538-3881/ac134b}

\bibitem[{Das {et~al.}(2021)Das, Rao, \& Yang}]{das21}
Das, S., Rao, S.~S., \& Yang, J. 2021, Journal of Time Series Analysis,
  \dodoi{10.1111/jtsa.12584}

\bibitem[{{Desort} {et~al.}(2007){Desort}, {Lagrange}, {Galland}, {Udry}, \&
  {Mayor}}]{desort07}
{Desort}, M., {Lagrange}, A.~M., {Galland}, F., {Udry}, S., \& {Mayor}, M.
  2007, \aap, 473, 983, \dodoi{10.1051/0004-6361:20078144}

\bibitem[{{DeWarf} {et~al.}(2010){DeWarf}, {Datin}, \& {Guinan}}]{dewarf10}
{DeWarf}, L.~E., {Datin}, K.~M., \& {Guinan}, E.~F. 2010, \apj, 722, 343,
  \dodoi{10.1088/0004-637X/722/1/343}

\bibitem[{Dodson-Robinson(2022)}]{dodsonrobinson22}
Dodson-Robinson, S. 2022, {NWelch: Spectral analysis of time series with
  nonuniform observing cadence}, 1.0.0,  Zenodo, \dodoi{10.5281/zenodo.5903196}

\bibitem[{{Dumusque} {et~al.}(2012){Dumusque}, {Pepe}, {Lovis},
  {S{\'e}gransan}, {Sahlmann}, {Benz}, {Bouchy}, {Mayor}, {Queloz}, {Santos},
  \& {Udry}}]{dumusque12}
{Dumusque}, X., {Pepe}, F., {Lovis}, C., {et~al.} 2012, \nat, 491, 207,
  \dodoi{10.1038/nature11572}

\bibitem[{{Endl} {et~al.}(2016){Endl}, {Brugamyer}, {Cochran}, {MacQueen},
  {Robertson}, {Meschiari}, {Ramirez}, {Shetrone}, {Gullikson}, {Johnson},
  {Wittenmyer}, {Horner}, {Ciardi}, {Horch}, {Simon}, {Howell}, {Everett},
  {Caldwell}, \& {Castanheira}}]{endl16}
{Endl}, M., {Brugamyer}, E.~J., {Cochran}, W.~D., {et~al.} 2016, \apj, 818, 34,
  \dodoi{10.3847/0004-637X/818/1/34}

\bibitem[{Enochson \& Goodman(1965)}]{enochson65}
Enochson, L.~D., \& Goodman, N.~R. 1965, Gaussian approximations to the
  distribution of sample coherence, Tech. Rep. Technical Report AFFDL TR
  65–67, Research and Tech. Div., AFSC, Wright–Patterson Air Force Base,
  Ohio

\bibitem[{{Feng} {et~al.}(2016){Feng}, {Tuomi}, {Jones}, {Butler}, \&
  {Vogt}}]{feng16}
{Feng}, F., {Tuomi}, M., {Jones}, H.~R.~A., {Butler}, R.~P., \& {Vogt}, S.
  2016, \mnras, 461, 2440, \dodoi{10.1093/mnras/stw1478}

\bibitem[{Fisher(1929)}]{fisher29}
Fisher, R.~A. 1929, Metron, 1, 3

\bibitem[{{Forveille} {et~al.}(2011){Forveille}, {Bonfils}, {Delfosse},
  {Alonso}, {Udry}, {Bouchy}, {Gillon}, {Lovis}, {Neves}, {Mayor}, {Pepe},
  {Queloz}, {Santos}, {Segransan}, {Almenara}, {Deeg}, \&
  {Rabus}}]{forveille11}
{Forveille}, T., {Bonfils}, X., {Delfosse}, X., {et~al.} 2011, arXiv e-prints,
  arXiv:1109.2505.
\newblock \doarXiv{1109.2505}

\bibitem[{{Gilbert} {et~al.}(2021){Gilbert}, {Barclay}, {Quintana},
  {Walkowicz}, {Vega}, {Schlieder}, {Monsue}, {Cale}, {Collins}, {Gaidos}, {El
  Mufti}, {Reefe}, {Plavchan}, {Tanner}, {Wittenmyer}, {Wittrock}, {Jenkins},
  {Latham}, {Ricker}, {Rose}, {Seager}, {Vanderspek}, \& {Winn}}]{gilbert21}
{Gilbert}, E.~A., {Barclay}, T., {Quintana}, E.~V., {et~al.} 2021, arXiv
  e-prints, arXiv:2109.03924.
\newblock \doarXiv{2109.03924}

\bibitem[{Godin(1972)}]{godin72}
Godin, G. 1972, The Analysis of Tides (University of Toronto Press)

\bibitem[{{Gomes da Silva} {et~al.}(2011){Gomes da Silva}, {Santos}, {Bonfils},
  {Delfosse}, {Forveille}, \& {Udry}}]{gomesdasilva11}
{Gomes da Silva}, J., {Santos}, N.~C., {Bonfils}, X., {et~al.} 2011, \aap, 534,
  A30, \dodoi{10.1051/0004-6361/201116971}

\bibitem[{{Gonz{\'a}lez-{\'A}lvarez} {et~al.}(2021){Gonz{\'a}lez-{\'A}lvarez},
  {Petralia}, {Micela}, {Maldonado}, {Affer}, {Maggio}, {Covino}, {Damasso},
  {Lanza}, {Perger}, {Pinamonti}, {Poretti}, {Scandariato}, {Sozzetti},
  {Bignamini}, {Giacobbe}, {Leto}, {Pagano}, {Zanmar S{\'a}nchez},
  {Gonz{\'a}lez Hern{\'a}ndez}, {Rebolo}, {Ribas}, {Su{\'a}rez Mascare{\~n}o},
  \& {Toledo-Padr{\'o}n}}]{gonzalezalvarez21}
{Gonz{\'a}lez-{\'A}lvarez}, E., {Petralia}, A., {Micela}, G., {et~al.} 2021,
  \aap, 649, A157, \dodoi{10.1051/0004-6361/202140490}

\bibitem[{{Gonz{\'a}lez Hern{\'a}ndez} {et~al.}(2018){Gonz{\'a}lez
  Hern{\'a}ndez}, {Pepe}, {Molaro}, \& {Santos}}]{gonzalezhernandez18}
{Gonz{\'a}lez Hern{\'a}ndez}, J.~I., {Pepe}, F., {Molaro}, P., \& {Santos},
  N.~C. 2018, ESPRESSO on VLT: An Instrument for Exoplanet Research, ed. H.~J.
  {Deeg} \& J.~A. {Belmonte}, 157, \dodoi{10.1007/978-3-319-55333-7\_157}

\bibitem[{{Gregory}(2011)}]{gregory11}
{Gregory}, P.~C. 2011, \mnras, 415, 2523,
  \dodoi{10.1111/j.1365-2966.2011.18877.x}

\bibitem[{{G{\"u}nther} {et~al.}(2020){G{\"u}nther}, {Berardo}, {Ducrot},
  {Murray}, {Stassun}, {Olah}, {Bouma}, {Rappaport}, {Winn}, {Feinstein},
  {Matthews}, {Sebastian}, {Rackham}, {Seli}, {Triaud}, {Gillen}, {Levine},
  {Demory}, {Gillon}, {Queloz}, {Ricker}, {Vanderspek}, {Seager}, {Latham},
  {Jenkins}, {Brasseur}, {Col{\'o}n}, {Daylan}, {Delrez}, {Garcia},
  {Jayaraman}, {Jehin}, {Kristiansen}, {Kruijssen}, {Philmann Pedersen},
  {Pozuelos}, {Rodriguez}, {Wohler}, \& {Zhan}}]{gunther20}
{G{\"u}nther}, M.~N., {Berardo}, D.~A., {Ducrot}, E., {et~al.} 2020, arXiv
  e-prints, arXiv:2008.11681.
\newblock \doarXiv{2008.11681}

\bibitem[{{Gupta} {et~al.}(2021){Gupta}, {Wright}, {Robertson}, {Halverson},
  {Luhn}, {Roy}, {Mahadevan}, {Ford}, {Bender}, {Blake}, {Hearty}, {Kanodia},
  {Logsdon}, {McElwain}, {Monson}, {Ninan}, {Schwab}, {Stef{\'a}nsson}, \&
  {Terrien}}]{gupta21}
{Gupta}, A.~F., {Wright}, J.~T., {Robertson}, P., {et~al.} 2021, \aj, 161, 130,
  \dodoi{10.3847/1538-3881/abd79e}

\bibitem[{Hannan \& Nicholls(1977)}]{hannan77}
Hannan, E.~J., \& Nicholls, D.~F. 1977, Journal of the American Statistical
  Association, 72, 834

\bibitem[{Harris(1978)}]{harris78}
Harris, F.~J. 1978, Proceedings of the IEEE, 66, 51

\bibitem[{{Hatzes}(2002)}]{hatzes02}
{Hatzes}, A.~P. 2002, Astronomische Nachrichten, 323, 392,
  \dodoi{10.1002/1521-3994(200208)323:3/4<392::AID-ASNA392>3.0.CO;2-M}

\bibitem[{{Heng} \& {Vogt}(2011)}]{heng11}
{Heng}, K., \& {Vogt}, S.~S. 2011, \mnras, 415, 2145,
  \dodoi{10.1111/j.1365-2966.2011.18853.x}

\bibitem[{{Henry} {et~al.}(1996){Henry}, {Soderblom}, {Donahue}, \&
  {Baliunas}}]{henry96}
{Henry}, T.~J., {Soderblom}, D.~R., {Donahue}, R.~A., \& {Baliunas}, S.~L.
  1996, \aj, 111, 439, \dodoi{10.1086/117796}

\bibitem[{{Hu{\'e}lamo} {et~al.}(2008){Hu{\'e}lamo}, {Figueira}, {Bonfils},
  {Santos}, {Pepe}, {Gillon}, {Azevedo}, {Barman}, {Fern{\'a}ndez}, {di Folco},
  {Guenther}, {Lovis}, {Melo}, {Queloz}, \& {Udry}}]{huelamo08}
{Hu{\'e}lamo}, N., {Figueira}, P., {Bonfils}, X., {et~al.} 2008, \aap, 489, L9,
  \dodoi{10.1051/0004-6361:200810596}

\bibitem[{Jenkins \& Watts(1968)}]{jenkins68}
Jenkins, G.~M., \& Watts, D.~G. 1968, Spectral Analysis and its Applications
  (Holden-Day, San Francisco)

\bibitem[{{Jurgenson} {et~al.}(2016){Jurgenson}, {Fischer}, {McCracken},
  {Sawyer}, {Szymkowiak}, {Davis}, {Muller}, \& {Santoro}}]{jurgenson16}
{Jurgenson}, C., {Fischer}, D., {McCracken}, T., {et~al.} 2016, in Society of
  Photo-Optical Instrumentation Engineers (SPIE) Conference Series, Vol. 9908,
  Ground-based and Airborne Instrumentation for Astronomy VI, ed. C.~J.
  {Evans}, L.~{Simard}, \& H.~{Takami}, 99086T, \dodoi{10.1117/12.2233002}

\bibitem[{{Kane} {et~al.}(2016){Kane}, {Thirumalachari}, {Henry}, {Hinkel},
  {Jensen}, {Boyajian}, {Fischer}, {Howard}, {Isaacson}, \& {Wright}}]{kane16}
{Kane}, S.~R., {Thirumalachari}, B., {Henry}, G.~W., {et~al.} 2016, \apjl, 820,
  L5, \dodoi{10.3847/2041-8205/820/1/L5}

\bibitem[{Keiner {et~al.}(2009)Keiner, Kunis, \& Potts}]{keiner2009using}
Keiner, J., Kunis, S., \& Potts, D. 2009, ACM Transactions on Mathematical
  Software (TOMS), 36, 1

\bibitem[{Krug {et~al.}(2019)Krug, Baars, Hutchins, \& Marusic}]{krug19}
Krug, D., Baars, W.~J., Hutchins, N., \& Marusic, I. 2019, Boundary-Layer
  Meteorology, 172, 199, \dodoi{10.1007/s10546-019-00445-4}

\bibitem[{Leroy(2012)}]{leroy2012fast}
Leroy, B. 2012, \aap, 545, A50

\bibitem[{Littlefair {et~al.}(2016)Littlefair, Burningham, \&
  Helling}]{littlefair16}
Littlefair, S.~P., Burningham, B., \& Helling, C. 2016, Monthly Notices of the
  Royal Astronomical Society, 466, 4250, \dodoi{10.1093/mnras/stw3376}

\bibitem[{Lomb(1976)}]{lomb76}
Lomb, N.~R. 1976, \apjs, 39, 447

\bibitem[{{Lubin} {et~al.}(2021){Lubin}, {Robertson}, {Stefansson}, {Ninan},
  {Mahadevan}, {Endl}, {Ford}, {Wright}, {Beard}, {Bender}, {Cochran},
  {Diddams}, {Fredrick}, {Halverson}, {Kanodia}, {Metcalf}, {Ramsey}, {Roy},
  {Schwab}, \& {Terrien}}]{lubin21}
{Lubin}, J., {Robertson}, P., {Stefansson}, G., {et~al.} 2021, \aj, 162, 61,
  \dodoi{10.3847/1538-3881/ac0057}

\bibitem[{{Maldonado} {et~al.}(2021){Maldonado}, {Petralia}, {Damasso},
  {Pinamonti}, {Scandariato}, {Gonz{\'a}lez-{\'A}lvarez}, {Affer}, {Micela},
  {Lanza}, {Leto}, {Poretti}, {Sozzetti}, {Perger}, {Giacobbe}, {Zanmar
  S{\'a}nchez}, {Maggio}, {Gonz{\'a}lez Hern{\'a}ndez}, {Rebolo}, {Ribas},
  {Su{\'a}rez-Mascare{\~n}o}, {Toledo-Padr{\'o}n}, {Bignamini}, {Molinari},
  {Covino}, {Claudi}, {Desidera}, {Herrero}, {Morales}, {Pagano}, \&
  {Piotto}}]{maldonado21}
{Maldonado}, J., {Petralia}, A., {Damasso}, M., {et~al.} 2021, \aap, 651, A93,
  \dodoi{10.1051/0004-6361/202141141}

\bibitem[{{Mayor} {et~al.}(2009){Mayor}, {Bonfils}, {Forveille}, {Delfosse},
  {Udry}, {Bertaux}, {Beust}, {Bouchy}, {Lovis}, {Pepe}, {Perrier}, {Queloz},
  \& {Santos}}]{mayor09}
{Mayor}, M., {Bonfils}, X., {Forveille}, T., {et~al.} 2009, \aap, 507, 487,
  \dodoi{10.1051/0004-6361/200912172}

\bibitem[{Miller \& Kelley(2021)}]{miller21}
Miller, C.~A., \& Kelley, A.~L. 2021, Limnology and Oceanography, 66, 1475,
  \dodoi{https://doi.org/10.1002/lno.11698}

\bibitem[{{Mortier} {et~al.}(2015){Mortier}, {Faria}, {Correia}, {Santerne}, \&
  {Santos}}]{mortier15}
{Mortier}, A., {Faria}, J.~P., {Correia}, C.~M., {Santerne}, A., \& {Santos},
  N.~C. 2015, \aap, 573, A101, \dodoi{10.1051/0004-6361/201424908}

\bibitem[{{Newton} {et~al.}(2016){Newton}, {Irwin}, {Charbonneau},
  {Berta-Thompson}, \& {Dittmann}}]{newton16}
{Newton}, E.~R., {Irwin}, J., {Charbonneau}, D., {Berta-Thompson}, Z.~K., \&
  {Dittmann}, J.~A. 2016, \apjl, 821, L19, \dodoi{10.3847/2041-8205/821/1/L19}

\bibitem[{{Nordlund} {et~al.}(2009){Nordlund}, {Stein}, \&
  {Asplund}}]{nordlund09}
{Nordlund}, {\r{A}}., {Stein}, R.~F., \& {Asplund}, M. 2009, Living Reviews in
  Solar Physics, 6, 2, \dodoi{10.12942/lrsp-2009-2}

\bibitem[{{Noyes} {et~al.}(1984){Noyes}, {Hartmann}, {Baliunas}, {Duncan}, \&
  {Vaughan}}]{noyes84}
{Noyes}, R.~W., {Hartmann}, L.~W., {Baliunas}, S.~L., {Duncan}, D.~K., \&
  {Vaughan}, A.~H. 1984, \apj, 279, 763, \dodoi{10.1086/161945}

\bibitem[{{\'O}lafsd{\'o}ttir {et~al.}(2016){\'O}lafsd{\'o}ttir, Schulz, \&
  Mudelsee}]{olafsdottir16}
{\'O}lafsd{\'o}ttir, K.~B., Schulz, M., \& Mudelsee, M. 2016, Computers \&
  Geosciences, 91, 11

\bibitem[{Pardo-Ig\'{u}zquiza \& Rodr\'{i}guez-Tovar(2012)}]{pardoiguzquiza12}
Pardo-Ig\'{u}zquiza, E., \& Rodr\'{i}guez-Tovar, F.~J. 2012, Computers \&
  Geosciences, 49, 207, \dodoi{https://doi.org/10.1016/j.cageo.2012.06.018}

\bibitem[{{Pepe} {et~al.}(2000){Pepe}, {Mayor}, {Delabre}, {Kohler}, {Lacroix},
  {Queloz}, {Udry}, {Benz}, {Bertaux}, \& {Sivan}}]{pepe00}
{Pepe}, F., {Mayor}, M., {Delabre}, B., {et~al.} 2000, in Society of
  Photo-Optical Instrumentation Engineers (SPIE) Conference Series, Vol. 4008,
  Optical and IR Telescope Instrumentation and Detectors, ed. M.~{Iye} \& A.~F.
  {Moorwood}, 582--592, \dodoi{10.1117/12.395516}

\bibitem[{{Percival}(1994)}]{percival94}
{Percival}, D.~B. 1994, Methods of Experimental Physics, 28, 313,
  \dodoi{10.1016/S0076-695X(08)60261-6}

\bibitem[{Percival \& Walden(2020)}]{percival93}
Percival, D.~B., \& Walden, A.~T. 2020, Spectral Analysis for Physical
  Applications: Multitaper and Conventional Univariate Techniques (Cambridge
  University Press)

\bibitem[{{Perger} {et~al.}(2021){Perger}, {Anglada-Escud{\'e}}, {Ribas},
  {Rosich}, {Herrero}, \& {Morales}}]{perger21}
{Perger}, M., {Anglada-Escud{\'e}}, G., {Ribas}, I., {et~al.} 2021, \aap, 645,
  A58, \dodoi{10.1051/0004-6361/202039594}

\bibitem[{{Perger} {et~al.}(2017){Perger}, {Garc{\'\i}a-Piquer}, {Ribas},
  {Morales}, {Affer}, {Micela}, {Damasso}, {Su{\'a}rez-Mascare{\~n}o},
  {Gonz{\'a}lez-Hern{\'a}ndez}, {Rebolo}, {Herrero}, {Rosich}, {Lafarga},
  {Bignamini}, {Sozzetti}, {Claudi}, {Cosentino}, {Molinari}, {Maldonado},
  {Maggio}, {Lanza}, {Poretti}, {Pagano}, {Desidera}, {Gratton}, {Piotto},
  {Bonomo}, {Martinez Fiorenzano}, {Giacobbe}, {Malavolta}, {Nascimbeni},
  {Rainer}, \& {Scandariato}}]{perger17a}
{Perger}, M., {Garc{\'\i}a-Piquer}, A., {Ribas}, I., {et~al.} 2017, \aap, 598,
  A26, \dodoi{10.1051/0004-6361/201628985}

\bibitem[{{Perugini} {et~al.}(2021){Perugini}, {Marsden}, {Waite}, {Jeffers},
  {Piskunov}, {Shaw}, {Burton}, {Mengel}, {Hughes}, \&
  {H{\'e}brard}}]{perugini21}
{Perugini}, G.~M., {Marsden}, S.~C., {Waite}, I.~A., {et~al.} 2021, \mnras,
  508, 3304, \dodoi{10.1093/mnras/stab2711}

\bibitem[{{Pierrehumbert}(2011)}]{pierrehumbert11}
{Pierrehumbert}, R.~T. 2011, \apjl, 726, L8, \dodoi{10.1088/2041-8205/726/1/L8}

\bibitem[{Podesta(2006)}]{podesta06}
Podesta, J.~J. 2006, Journal of Geophysical Research: Space Physics, 111,
  \dodoi{https://doi.org/10.1029/2005JA011233}

\bibitem[{Press \& Rybicki(1989)}]{press1989fast}
Press, W.~H., \& Rybicki, G.~B. 1989, \apj, 338, 277

\bibitem[{Pukkila \& Nyquist(1985)}]{pukkila85}
Pukkila, T., \& Nyquist, H. 1985, Biometrika, 72, 317

\bibitem[{Queloz {et~al.}(2001)Queloz, Henry, Sivan, Baliunas, Beuzit, Donahue,
  Mayor, Naef, Perrier, \& Udry}]{queloz01}
Queloz, D., Henry, G., Sivan, J., {et~al.} 2001, \aap, 379, 279,
  \dodoi{doi:10.1051/0004-6361:20011308}

\bibitem[{{Queloz} {et~al.}(2009){Queloz}, {Bouchy}, {Moutou}, {Hatzes},
  {H{\'e}brard}, {Alonso}, {Auvergne}, {Baglin}, {Barbieri}, {Barge}, {Benz},
  {Bord{\'e}}, {Deeg}, {Deleuil}, {Dvorak}, {Erikson}, {Ferraz Mello},
  {Fridlund}, {Gandolfi}, {Gillon}, {Guenther}, {Guillot}, {Jorda}, {Hartmann},
  {Lammer}, {L{\'e}ger}, {Llebaria}, {Lovis}, {Magain}, {Mayor}, {Mazeh},
  {Ollivier}, {P{\"a}tzold}, {Pepe}, {Rauer}, {Rouan}, {Schneider},
  {Segransan}, {Udry}, \& {Wuchterl}}]{queloz09}
{Queloz}, D., {Bouchy}, F., {Moutou}, C., {et~al.} 2009, \aap, 506, 303,
  \dodoi{10.1051/0004-6361/200913096}

\bibitem[{{Rajpaul} {et~al.}(2016){Rajpaul}, {Aigrain}, \&
  {Roberts}}]{rajpaul16}
{Rajpaul}, V., {Aigrain}, S., \& {Roberts}, S. 2016, \mnras, 456, L6,
  \dodoi{10.1093/mnrasl/slv164}

\bibitem[{{Rajpaul} {et~al.}(2021){Rajpaul}, {Buchhave}, {Lacedelli}, {Rice},
  {Mortier}, {Malavolta}, {Aigrain}, {Borsato}, {Mayo}, {Charbonneau},
  {Damasso}, {Dumusque}, {Ghedina}, {Latham}, {L{\'o}pez-Morales},
  {Magazz{\'u}}, {Micela}, {Molinari}, {Pepe}, {Piotto}, {Poretti}, {Rowther},
  {Sozzetti}, {Udry}, \& {Watson}}]{rajpaul21}
{Rajpaul}, V.~M., {Buchhave}, L.~A., {Lacedelli}, G., {et~al.} 2021, \mnras,
  \dodoi{10.1093/mnras/stab2192}

\bibitem[{{Reinhold} {et~al.}(2013){Reinhold}, {Reiners}, \&
  {Basri}}]{reinhold13}
{Reinhold}, T., {Reiners}, A., \& {Basri}, G. 2013, \aap, 560, A4,
  \dodoi{10.1051/0004-6361/201321970}

\bibitem[{{Robertson} {et~al.}(2013){Robertson}, {Endl}, {Cochran}, \&
  {Dodson-Robinson}}]{robertson13}
{Robertson}, P., {Endl}, M., {Cochran}, W.~D., \& {Dodson-Robinson}, S.~E.
  2013, \apj, 764, 3, \dodoi{10.1088/0004-637X/764/1/3}

\bibitem[{{Robertson} \& {Mahadevan}(2014)}]{robertson14a}
{Robertson}, P., \& {Mahadevan}, S. 2014, \apjl, 793, L24,
  \dodoi{10.1088/2041-8205/793/2/L24}

\bibitem[{{Robertson} {et~al.}(2014){Robertson}, {Mahadevan}, {Endl}, \&
  {Roy}}]{robertson14b}
{Robertson}, P., {Mahadevan}, S., {Endl}, M., \& {Roy}, A. 2014, Science, 345,
  440, \dodoi{10.1126/science.1253253}

\bibitem[{Robertson {et~al.}(2015)Robertson, Roy, \& Mahadevan}]{robertson15}
Robertson, P., Roy, A., \& Mahadevan, S. 2015, \apjl, 805, L22,
  \dodoi{doi:10.1088/2041-8205/805/2/L22}

\bibitem[{{Rodono} {et~al.}(1986){Rodono}, {Cutispoto}, {Pazzani}, {Catalano},
  {Byrne}, {Doyle}, {Butler}, {Andrews}, {Blanco}, {Marilli}, {Linsky},
  {Scaltriti}, {Busso}, {Cellino}, {Hopkins}, {Okazaki}, {Hayashi}, {Zeilik},
  {Helston}, {Henson}, {Smith}, \& {Simon}}]{rodono86}
{Rodono}, M., {Cutispoto}, G., {Pazzani}, V., {et~al.} 1986, \aap, 165, 135

\bibitem[{{Saar} \& {Donahue}(1997)}]{saar97}
{Saar}, S.~H., \& {Donahue}, R.~A. 1997, \apj, 485, 319, \dodoi{10.1086/304392}

\bibitem[{{Sarkis} {et~al.}(2018){Sarkis}, {Henning}, {K{\"u}rster},
  {Trifonov}, {Zechmeister}, {Tal-Or}, {Anglada-Escud{\'e}}, {Hatzes},
  {Lafarga}, {Dreizler}, {Ribas}, {Caballero}, {Reiners}, {Mallonn}, {Morales},
  {Kaminski}, {Aceituno}, {Amado}, {B{\'e}jar}, {Hagen}, {Jeffers},
  {Quirrenbach}, {Launhardt}, {Marvin}, \& {Montes}}]{sarkis18}
{Sarkis}, P., {Henning}, T., {K{\"u}rster}, M., {et~al.} 2018, \aj, 155, 257,
  \dodoi{10.3847/1538-3881/aac108}

\bibitem[{Scafetta \& Mazzarella(2015)}]{scafetta15}
Scafetta, N., \& Mazzarella, A. 2015, Natural Hazards, 76, 1807.
\newblock \url{https://doi.org/10.1007/s11069-014-1571-z}

\bibitem[{Scargle(1982)}]{scargle82}
Scargle, J.~D. 1982, \apj, 263, 835

\bibitem[{Scargle(1989)}]{scargle89}
---. 1989, \apj, 343, 874

\bibitem[{Schulz \& Stattegger(1997)}]{schulz97}
Schulz, M., \& Stattegger, K. 1997, Computers \& Geosciences, 23, 929,
  \dodoi{10.1016/S0098-3004(97)00087-3}

\bibitem[{Schuster(1898)}]{schuster1898}
Schuster, A. 1898, Terrestrial Magnetism and Atmospheric Electricity, 3, 13

\bibitem[{{Shkolnik} {et~al.}(2003){Shkolnik}, {Walker}, \&
  {Bohlender}}]{shkolnik03}
{Shkolnik}, E., {Walker}, G.~A.~H., \& {Bohlender}, D.~A. 2003, \apj, 597,
  1092, \dodoi{10.1086/378583}

\bibitem[{Shumway \& Stoffer(2001)}]{shumwaystoffer}
Shumway, R.~H., \& Stoffer, D.~S. 2001, Time Series Analysis and Its
  Applications, 4th edn. (New York: Springer)

\bibitem[{Siegel(1980)}]{siegel80}
Siegel, A. 1980, Journal of the American Statistical Association, 75, 345

\bibitem[{Slepian(1978)}]{s78}
Slepian, D. 1978, Bell System Technical Journal, 57, 1371

\bibitem[{{Springford} {et~al.}(2020){Springford}, {Eadie}, \&
  {Thomson}}]{springford2020improving}
{Springford}, A., {Eadie}, G.~M., \& {Thomson}, D.~J. 2020, \aj, 159, 205,
  \dodoi{10.3847/1538-3881/ab7fa1}

\bibitem[{Stoica {et~al.}(1989)Stoica, Moses, Friedlander, \&
  S{\"o}derstr{\"o}m}]{stoica89}
Stoica, P., Moses, R.~L., Friedlander, B., \& S{\"o}derstr{\"o}m, T. 1989, IEEE
  Trans. Acoust. Speech Signal Process., 37, 378

\bibitem[{{Su{\'a}rez Mascare{\~n}o} {et~al.}(2017){Su{\'a}rez Mascare{\~n}o},
  {Rebolo}, {Gonz{\'a}lez Hern{\'a}ndez}, \& {Esposito}}]{suarezmascareno17b}
{Su{\'a}rez Mascare{\~n}o}, A., {Rebolo}, R., {Gonz{\'a}lez Hern{\'a}ndez},
  J.~I., \& {Esposito}, M. 2017, \mnras, 468, 4772,
  \dodoi{10.1093/mnras/stx771}

\bibitem[{{Tal-Or} {et~al.}(2018){Tal-Or}, {Zechmeister}, {Reiners}, {Jeffers},
  {Sch{\"o}fer}, {Quirrenbach}, {Amado}, {Ribas}, {Caballero}, {Aceituno},
  {Bauer}, {B{\'e}jar}, {Czesla}, {Dreizler}, {Fuhrmeister}, {Hatzes},
  {Johnson}, {K{\"u}rster}, {Lafarga}, {Montes}, {Morales}, {Reffert},
  {Sadegi}, {Seifert}, \& {Shulyak}}]{talor18}
{Tal-Or}, L., {Zechmeister}, M., {Reiners}, A., {et~al.} 2018, \aap, 614, A122,
  \dodoi{10.1051/0004-6361/201732362}

\bibitem[{Thompson {et~al.}(2017)Thompson, Watson, de~Mooij, \&
  Jess}]{thompson17}
Thompson, A., Watson, C., de~Mooij, E., \& Jess, D. 2017, Monthly Notices of
  the Royal Astronomical Society: Letters, 468, L16,
  \dodoi{doi:10.1093/mnrasl/slx018}

\bibitem[{Thomson(1982)}]{t82}
Thomson, D.~J. 1982, Proceedings of the IEEE, 70, 1055

\bibitem[{Thomson(1994)}]{thomson94}
Thomson, D.~J. 1994, in Proceedings of ICASSP'94. IEEE International Conference
  on Acoustics, Speech and Signal Processing, Vol.~6, IEEE, VI--73

\bibitem[{{Thomson}(1995)}]{thomson95}
{Thomson}, D.~J. 1995, Science, 268, 59, \dodoi{10.1126/science.268.5207.59}

\bibitem[{Thomson \& Chave(1991)}]{tc91}
Thomson, D.~J., \& Chave, A.~D. 1991, in Advances in Spectrum Analysis and
  Array Processing, ed. S.~Haykin, Vol.~1 (Upper Saddle River, NJ:
  Prentice-Hall), 58--113

\bibitem[{Thomson \& Haley(2014)}]{th14}
Thomson, D.~J., \& Haley, C.~L. 2014, Proceedings of the Royal Society A:
  Mathematical, Physical and Engineering Sciences, 470, 20140101,
  \dodoi{10.1098/rspa.2014.0101}

\bibitem[{Toledo-Padr{\'o}n {et~al.}(2019)Toledo-Padr{\'o}n,
  Gonz{\'a}lez~Hern{\'a}ndez, Rodr{\'\i}guez-L{\'o}pez,
  Su{\'a}rez~Mascare{\~n}o, Rebolo, Butler, Ribas, Anglada-Escud{\'e}, Johnson,
  Reiners, {et~al.}}]{toledopadron19}
Toledo-Padr{\'o}n, B., Gonz{\'a}lez~Hern{\'a}ndez, J.,
  Rodr{\'\i}guez-L{\'o}pez, C., {et~al.} 2019, Monthly Notices of the Royal
  Astronomical Society, 488, 5145

\bibitem[{{Toner} \& {Gray}(1988)}]{toner88}
{Toner}, C.~G., \& {Gray}, D.~F. 1988, \apj, 334, 1008, \dodoi{10.1086/166893}

\bibitem[{{Tuomi} {et~al.}(2013){Tuomi}, {Jones}, {Jenkins}, {Tinney},
  {Butler}, {Vogt}, {Barnes}, {Wittenmyer}, {O'Toole}, {Horner}, {Bailey},
  {Carter}, {Wright}, {Salter}, \& {Pinfield}}]{tuomi13}
{Tuomi}, M., {Jones}, H.~R.~A., {Jenkins}, J.~S., {et~al.} 2013, \aap, 551,
  A79, \dodoi{10.1051/0004-6361/201220509}

\bibitem[{{Udry} {et~al.}(2007){Udry}, {Bonfils}, {Delfosse}, {Forveille},
  {Mayor}, {Perrier}, {Bouchy}, {Lovis}, {Pepe}, {Queloz}, \&
  {Bertaux}}]{udry07}
{Udry}, S., {Bonfils}, X., {Delfosse}, X., {et~al.} 2007, \aap, 469, L43,
  \dodoi{10.1051/0004-6361:20077612}

\bibitem[{{Vanderburg} {et~al.}(2016){Vanderburg}, {Plavchan}, {Johnson},
  {Ciardi}, {Swift}, \& {Kane}}]{vanderburg16}
{Vanderburg}, A., {Plavchan}, P., {Johnson}, J.~A., {et~al.} 2016, \mnras, 459,
  3565, \dodoi{10.1093/mnras/stw863}

\bibitem[{{Vogt} {et~al.}(2010){Vogt}, {Butler}, {Rivera}, {Haghighipour},
  {Henry}, \& {Williamson}}]{vogt10}
{Vogt}, S.~S., {Butler}, R.~P., {Rivera}, E.~J., {et~al.} 2010, \apj, 723, 954,
  \dodoi{10.1088/0004-637X/723/1/954}

\bibitem[{{von Bloh} {et~al.}(2011){von Bloh}, {Cuntz}, {Franck}, \&
  {Bounama}}]{vonbloh11}
{von Bloh}, W., {Cuntz}, M., {Franck}, S., \& {Bounama}, C. 2011, \aap, 528,
  A133, \dodoi{10.1051/0004-6361/201116534}

\bibitem[{{Walker} \& {Stephenson}(2014)}]{walker14}
{Walker}, A.~D.~M., \& {Stephenson}, J.~A.~E. 2014, Annales Geophysicae, 32,
  1217, \dodoi{10.5194/angeo-32-1217-2014}

\bibitem[{Welch(1967)}]{welch67}
Welch, P. 1967, IEEE Transactions on Audio and Electroacoustics, AU-15, 70

\bibitem[{{Wise} {et~al.}(2018){Wise}, {Dodson-Robinson}, {Bevenour}, \&
  {Provini}}]{wise18}
{Wise}, A.~W., {Dodson-Robinson}, S.~E., {Bevenour}, K., \& {Provini}, A. 2018,
  \aj, 156, 180, \dodoi{10.3847/1538-3881/aadd94}

\bibitem[{{Wordsworth} {et~al.}(2011){Wordsworth}, {Forget}, {Selsis},
  {Millour}, {Charnay}, \& {Madeleine}}]{wordsworth11}
{Wordsworth}, R.~D., {Forget}, F., {Selsis}, F., {et~al.} 2011, \apjl, 733,
  L48, \dodoi{10.1088/2041-8205/733/2/L48}

\bibitem[{{Zechmeister} \& {K{\"u}rster}(2009)}]{zechmeister09}
{Zechmeister}, M., \& {K{\"u}rster}, M. 2009, \aap, 496, 577,
  \dodoi{10.1051/0004-6361:200811296}

\end{thebibliography}




\end{document}